\documentclass[useAMS,usenatbib]{mn2e}
\usepackage{amssymb,amsmath}

\usepackage{graphicx}
\usepackage{natbib}
\usepackage{float}
\usepackage{url}

\title[Li in T-Tauri stars]{No evidence for intense, cold accretion onto YSOs from measurements of Li in T-Tauri stars}
\author[Sergison et al.]{Darryl J. Sergison,$^{1}$\thanks{E-mail:
darryl@astro.ex.ac.uk}  N. J. Mayne$^{1}$, Tim Naylor$^{1}$,
R. D. Jeffries$^{2}$ \& Cameron P. M. Bell$^{1}$\\
$^{1}$School of Physics, University of Exeter, Stocker Road, Exeter
EX4 4QL\\
$^{2}$Astrophysics Group, Keele University, Keele, Staffordshire ST5 5BG}

\begin{document}

\date{Accepted date}

\pagerange{\pageref{firstpage}--\pageref{lastpage}} \pubyear{2002}

\maketitle

\label{firstpage}

\begin{abstract}
We have used medium resolution spectra to search for evidence that proto-stellar objects accrete at
high rates during their early `assembly phase'. Models
predict that depleted lithium and reduced luminosity in T-Tauri stars are key signatures of
`cold' high-rate accretion occurring early in a star's evolution.

We found no evidence in 168 stars in NGC 2264 and the
Orion Nebula Cluster for
strong lithium depletion through analysis of veiling
corrected 6708$\,$\AA\ lithium spectral line strengths. This suggests
that `cold' accretion at high rates  ($\dot{M}\ge5\times10^{-4}$
M$_{\rm{\odot}}$ yr$^{-1}$)
occurs in the assembly phase of fewer
than 0.5 per cent of $0.3 \le M_{\star} \le 1.9\,\rm{M}_{\odot}$ stars.

We also find that the dispersion in the strength of
the 6708$\,$\AA\ lithium line might imply an age spread that is
similar in magnitude to the apparent age spread implied by the
luminosity dispersion seen in colour magnitude diagrams. Evidence for weak lithium
depletion ($<10$ per cent in equivalent width) that
is correlated with luminosity is also apparent, but we are unable to determine whether age spreads or
accretion at rates less than $5\times10^{-4}$
M$_{\rm{\odot}}$ yr$^{-1}$ are responsible.

\end{abstract}


\section{Introduction}

The radius and luminosity of a pre-main-sequence (pre-MS) star 
are determined by the balance between its Kelvin-Helmholtz (K-H) contraction
timescale and accretion activity. If accretion increases the mass of
the star more rapidly than the K-H timescale allows energy to be radiated, it will
displace the star from thermal equilibrium and cause it to exhibit a
radius and luminosity that is inconsistent with non-accreting models
\citep{1999MNRAS.310..360T,2009ApJ...702L..27B,2011arXiv1106.3343H}. As the age of a T-Tauri star ($\sim 2-5\,\rm{Myr}$) can be less than the K-H timescale,
its radius and luminosity may still exhibit the influence of its prior
accretion history in the embedded phase.

The core conditions of an embedded proto-star are potentially dependent on its
mass accretion rate. Accretion of material with low internal energy
(`cold' accretion), at rates in excess of
$\dot{M}\ge10^{-4}\,\rm{M}_{\odot}\,\rm{yr}^{-1}$ may gravitationally
compress the star, increasing its core temperature and triggering the
early onset of lithium burning \citep{2010A&A...521A..44B}. Efficient large-scale convection (seen in pre-MS stars) would then
rapidly deplete lithium throughout the star. Observationally, this
mechanism may be apparent in the later T-Tauri phase as a surface
lithium abundance that is much lower than that expected for a given
age. In order for lithium to be depleted, a star must be radially
compressed and hence the luminosity must be affected. As a result, lithium
abundance in combination with luminosity 
may be a powerful observational signature of heavy accretion during the early
evolution of pre-MS stars.
\begin{figure*}
\includegraphics[height=6.5cm]{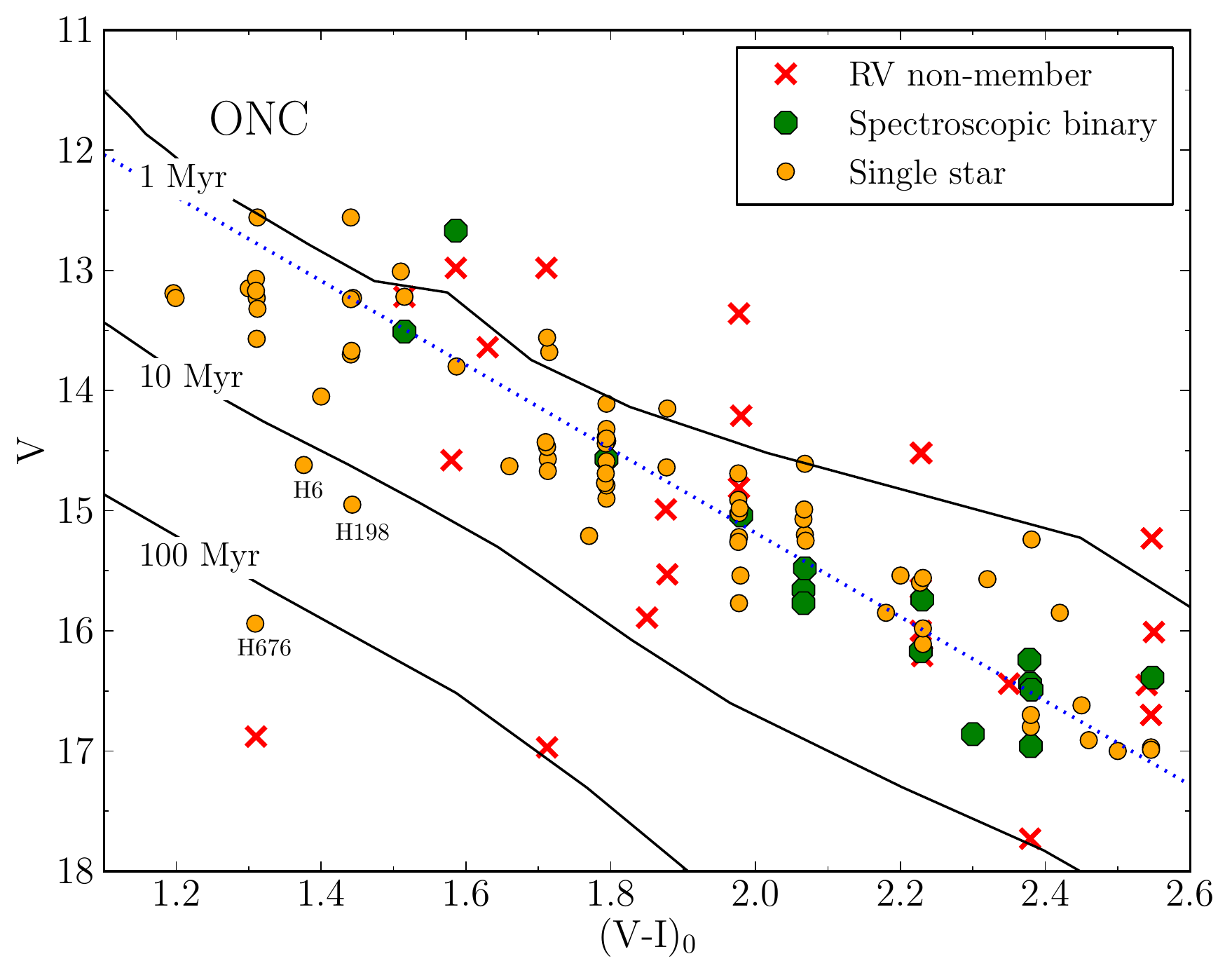}
\hspace{0.5cm}
\includegraphics[height=6.5cm]{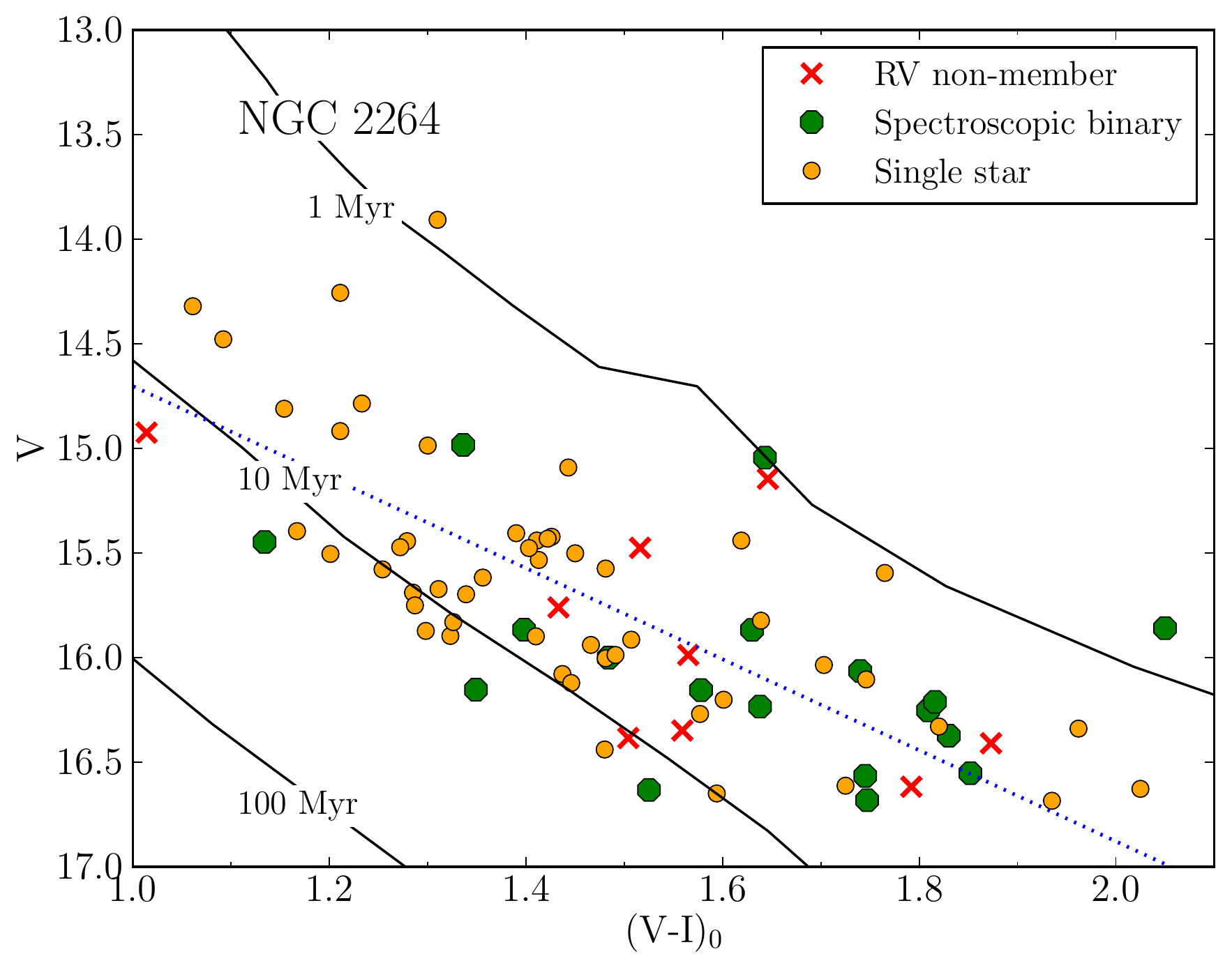}

 \caption{Colour-magnitude diagrams of target cluster members in the Orion Nebula Cluster (left)
   and NGC 2264 (right) compared with isochrones from
   \protect\cite{2000A&A...358..593S}. Medium orange (light grey) filled circles
   are confirmed single star members for which we have established veiling and EW[Li].
   Large green (dark grey) filled circles are spectroscopic
   binary stars (and confirmed members) for which we have EW[Li] lower limits. Red crosses
   depict stars that we observed but were identified as radial
   velocity non-members. The limits of the plot represent the
   photometric limits imposed on our sample by (i) the sensitivity
   of the observations (ii) the $T_{\rm{eff}}$ cuts applied for
   the mass range of interest.  Dotted (blue) line indicates the
   division between bright and faint samples defined in Section 5.2. Note: The three members that appear
   fainter than the 10 Myr isochrone in the ONC are numbered, these are discussed
   in Section 5.5. ONC photometry from \protect\cite{1997AJ....113.1733H}, NGC 2264 photometry from \protect\cite{2007MNRAS.375.1220M}. }
\end{figure*}

Current support for past accretion at high rates in early stellar evolution stems from three key observations.

\begin{enumerate}

\item \cite{1990AJ.....99..869K} find observational evidence for
  accretion at high rates contributing a significant fraction
of a star's mass during the early `assembly phase' of star
formation. Angular momentum conservation indicates that infall must proceed first
onto a circumstellar disc. Some of this material may then fall onto
the star in bursts, perhaps through gravitational instability. Models by \cite{2005ApJ...633L.137V} predict
heavy bouts ($\dot{M}\ge10^{-4}\,\rm{M}_{\odot}\,\rm{yr}^{-1}$) of
accretion lasting $\le\,100$ years, between quiescent periods
($\dot{M}\leq10^{-6}\,\rm{M}_{\odot}\,\rm{yr}^{-1}$) lasting
$1\--3\,$kyr. These timescales fit with current
observations of YSOs, explaining both the large population of
low-luminosity class I sources and the small fraction of very luminous
sources \citep{2009ApJ...692..973E}.\\

\item Many observations of both star-forming regions \citep[see][for a review]{2000ApJ...530..277E} and local molecular
clouds \citep{2007RMxAA..43..123B} appear to support the idea that star
formation occurs on the order of one to a few dynamical crossing times, which is
no more than a few Myr for a cluster such as the Orion Nebula Cluster (ONC)
\cite[e.g.][]{2006ApJ...641L.121T,2011MNRAS.418.1948J}. If this is
true then the colour-magnitude diagram (CMD) luminosity dispersion observed in young clusters  \citep[see Figure 1 and][] {1976AJ.....81..845M,2000ApJ...540..255P,2010ApJ...723..166D},
that is sometimes attributed to an age spread  $\geq 10$ Myr, may
simply be evidence of
variation in radius (and potentially early accretion history) within a
coeval population \citep{2007MNRAS.381.1169J}.\\

\item Stars that are unusually depleted in lithium have been reported
  in the young clusters $\sigma$ Ori and the ONC
  \citep{2005ApJ...626L..49P,2007ApJ...659L..41P,2007A&A...462L..23S}. Some
  of these lithium depleted objects are also faint
  compared with the main cluster population. These characteristics are
  explained by invoking ages that are at least 10 Myr older than the main
  cluster population. However, past accretion at high rates
  may also be a consistent explanation and does not need to invoke
  an age spread that is at odds with observations of star formation on a
  dynamical crossing time.\\

\end{enumerate}

Accreting material adds gravitational and thermal energy to the
stellar interior at a
rate defined by \cite{1997ApJ...475..770H} as

\begin{equation}
\frac{dE_{acc}}{dt} = \alpha\,\epsilon\,\frac{G\,M_{\star}\,\dot{M}}{R_{\star}}
\end{equation}
where $\epsilon \leq 1$ for material falling
from within the star's potential well and
$\epsilon \leq 0.5$ for material accreted through a boundary layer from a thin
disc. $\alpha$ is the fraction of the accretion luminosity absorbed
by the star. \cite{1997ApJ...475..770H} suggest that
$\alpha\ll\,1$, providing the fraction of the stellar surface that
accretion occurs over is small.

These qualitative ideas are backed by the results of numerical simulations, such
as those by \cite{1999MNRAS.310..360T}, \cite{2009ApJ...702L..27B}, \cite{2010A&A...521A..44B}, 
\cite{2011ApJ...738..140H} and \cite{2012ApJ...756..118B}.
\cite{2010A&A...521A..44B}  find that in the limit where $\alpha
\lesssim 0.2$ (termed `cold' accretion)
additional mass added to the star causes gravitational
contraction, reducing overall luminosity and increasing core
temperature and pressure. They also find that this process
significantly depletes lithium
abundance in very
young stars (from $\lesssim\,1\,\rm{Myr}$) with
$M_{\star}<2\,\rm{M}_{\odot}$. The models suggest that a threshold accretion rate exists
whereby detectable lithium depletion occurs if the accretion rate during bursts exceeds
$\dot{M}\ge 5\times10^{-4}\,\rm{M}_{\odot}\,\rm{yr}^{-1} $ and the
initial core mass of the proto-star has a mass $\leq\,0.03\,\rm{M}_{\odot}$ (see Figure 2). For a $1\,\rm{M}_{\odot}$ star, cold accretion
that is able to measurably change a star's luminosity is found to always cause a
measurable depletion in lithium.

Lithium depletion occurs as the abundance of lithium in the photosphere is a sensitive function of the temperature at the core of a fully convective star. Once temperatures exceed $\sim2.5\,\times\,10^6\,\rm{K}$
(slightly below that required for hydrogen burning), lithium is swiftly
depleted via  $^{7}\rm{Li}+\rm{p}\,\rightarrow\,^{4}\rm{He}\,+\alpha$
reactions. 

In contrast, non-accreting models suggest that lithium depletion will occur much later in the
star's evolution. \cite{2000A&A...358..593S} and
\cite{2002A&A...382..563B} predict less than 20 per cent lithium depletion for
stars of age $<10\,\rm{Myr}$ at masses above $0.6\,\rm{M}_{\rm{\odot}}$.

\begin{figure}
\begin{center}
\includegraphics[width=\columnwidth]{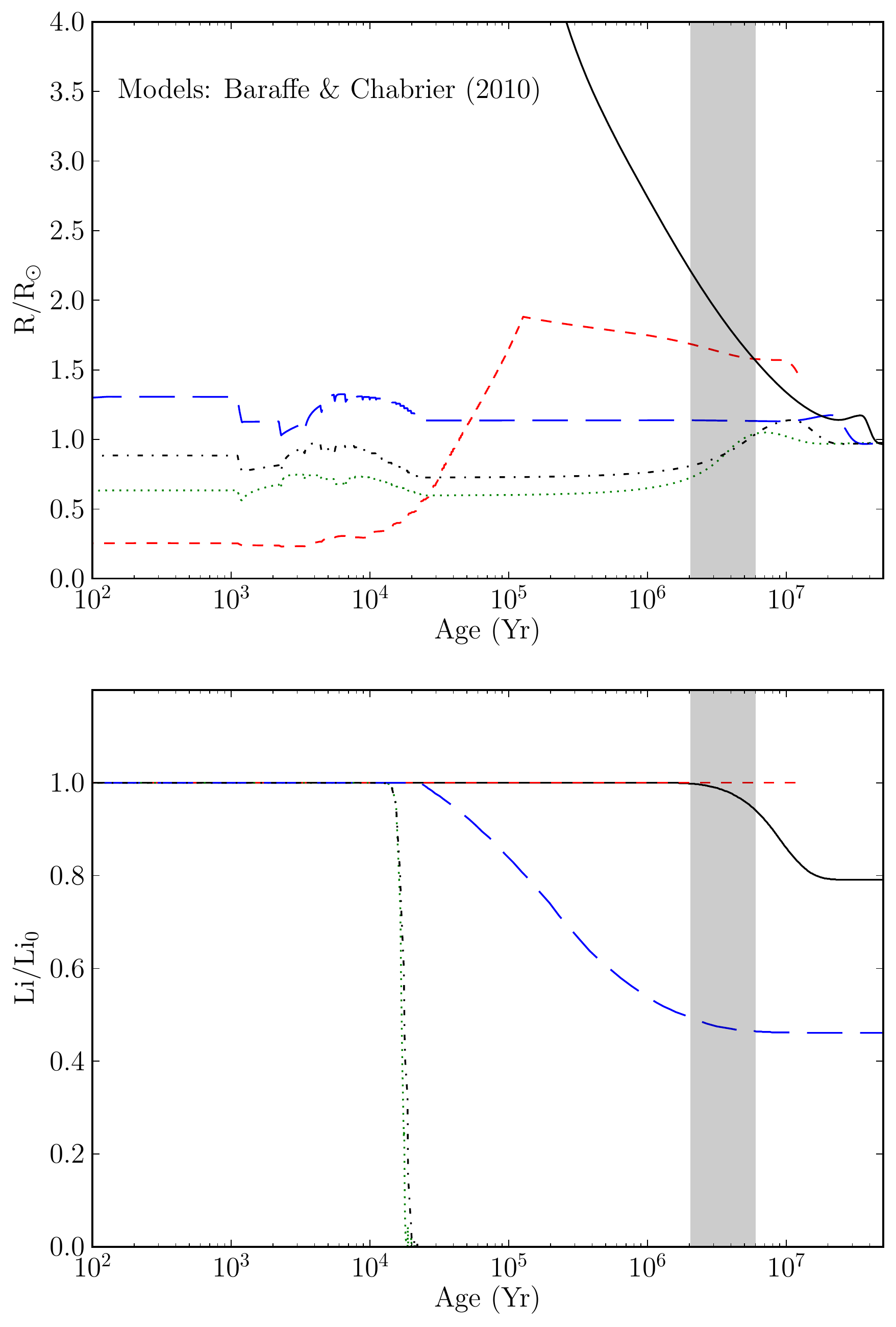}
\end{center}
\caption{Evolution of radius (upper panel) and photospheric lithium abundance
  (lower panel) as a function of
time for \protect\cite{2010A&A...521A..44B} models with episodic accretion and reaching a final mass of
$1\,\rm{M}_{\odot}$. Short dash (red): $M_{\rm{init}} = 1\,$M$_{\rm{jup}},
\dot{M}_{\rm{burst}} = 1 \times 10^{-4}\, \rm{M}_{\rm{\odot}} \rm{yr}^{-1},
N_{\rm{burst}} =  100$ ; dot (green): $M_{\rm{init}} = 10\,$M$_{\rm{jup}},
\dot{M}_{\rm{burst}} = 5 \times 10^{-4} \, \rm{M}_{\rm{\odot}} \rm{yr}^{-1},
N_{\rm{burst}} =  20$; dash-dot (black): $M_{\rm{init}} = 20\,$M$_{\rm{jup}},
\dot{M}_{\rm{burst}} = 5 \times 10^{-4} \, \rm{M}_{\rm{\odot}} \rm{yr}^{-1},
N_{\rm{burst}} =  20$; long dash (blue): $M_{\rm{init}} = 30\,$M$_{\rm{jup}},
\dot{M}_{\rm{burst}} = 5 \times 10^{-4} \, \rm{M}_{\rm{\odot}} \rm{yr}^{-1},
N_{\rm{burst}} =  20$; solid line (black) is a non-accreting model for
a $1\,\rm{M}_{\rm{\odot}}$ star. All calculations are made with $\Delta \rm{t}_{\rm{burst}}$
= 100 yr and $\Delta \rm{t}_{\rm{quiet}}$ = 1000 yr. The grey shaded region encompasses age
estimates for the clusters studied in this paper}
\end{figure}

Cold accretion models make two clear predictions which we test by observation.

\begin{enumerate}
\item If accretion occurs at $\dot{M} \ge 5 \times
10^{-4}\,\rm{M}_{\odot}\,\rm{yr}^{-1}$, some highly lithium depleted
stars should be found that cannot be explained by non-accreting
evolutionary models.\\
\item  In the cold accretion model, lithium depletion and
low luminosity result from the same mechanism, hence observed lithium depletion should
correlate with the appearance of old age in the CMD.\\

\end{enumerate}

 Section 2 describes the
theoretical models and the observation target selection. Section 3 details our
observations and data reduction process.  Section 4
presents our results, including radial velocity, binarity, accretion veiling,
$v\,$sin$\,i$, and lithium equivalent width. In Section 5 we discuss
the lithium and veiling results for each cluster and show that there
is some
correlation between lithium abundance and CMD position. In Section 6 we
conclude that there is no evidence for past episodes of cold accretion at very high rates.

\section[]{Theory and target selection}

\subsection{Theoretical lithium depletion}
Figure 3 shows simulated lithium isochrones based on several different
interior models \citep{2002A&A...382..563B,2000A&A...358..593S} and illustrates the predicted equivalent width of the
$6708\,\rm{\AA}$ lithium line (EW[Li]) as a function
 of $(V-I_{\rm{c}})_{0}$ colour. The isochrones were derived using a
colour -- $T_{\rm{eff}}$ relation defined by observations of
main-sequence stars
\citep{1995ApJS..101..117K}, curves of growth from
\cite{2003MNRAS.343.1271J} and an assumed initial lithium abundance of
A(Li) = 3.3 on the logarithmic scale of
\cite{1989GeCoA..53..197A}.
Comparison of the isochrones with those created using a Pleiades
tuned colour -- $T_{\rm eff}$ relation
\citep[from][]{2005MNRAS.358...13J} show that whilst the colour bounds of the
lithium depletion `chasm' vary as a
function of model, the location of the isochrones do
not vary by more than  $\pm 0.1$ in $(V-I_{\rm{c}})_{0}$.

\begin{figure*}
\begin{center}
\includegraphics[width=\textwidth]{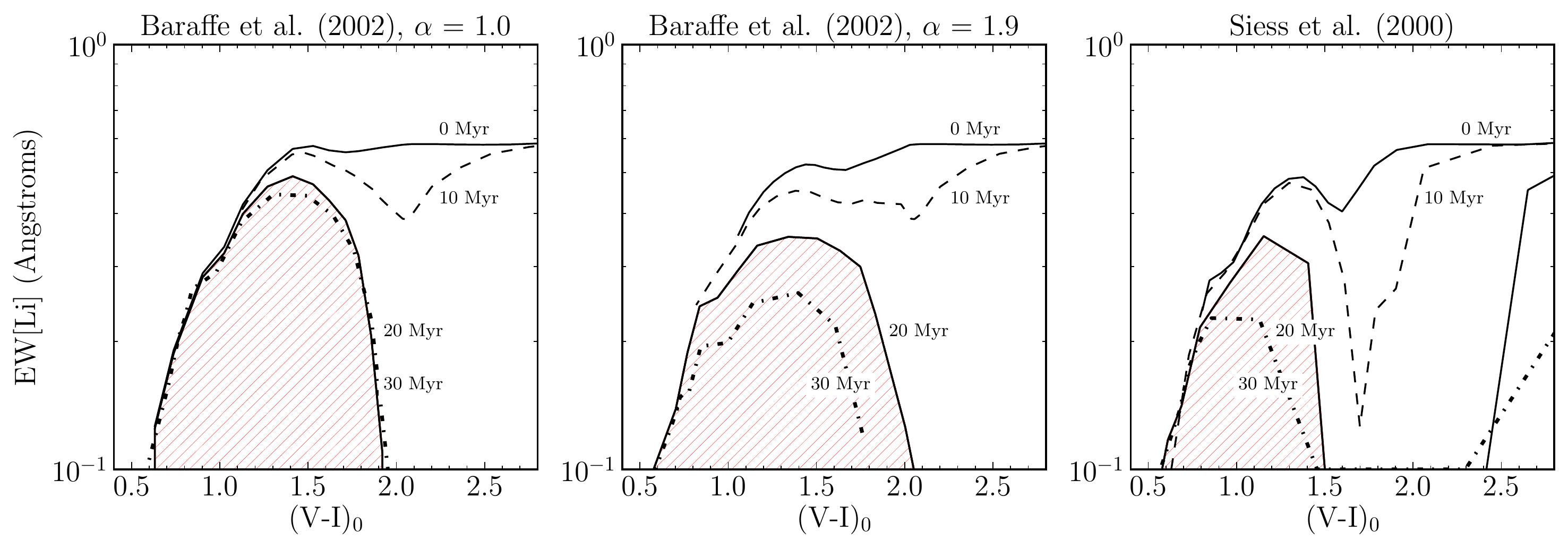}
\end{center}
\caption{Predicted evolution of the 6708\,\AA\ lithium line strength as a function
  of age and colour in non-accreting models of pre-MS stars. The (red) hatched region is
  where we expect to find no stars that are younger than 20 Myr, thus
  defining a `forbidden zone' in our clusters based on current age
  estimates. Stellar interior models are from
  \protect\cite{2002A&A...382..563B} and \protect\cite{2000A&A...358..593S},
  $T_{\rm{eff}}$ to $(V-I_{\rm{c}})_{0}$ from
  \protect\cite{1995ApJS..101..117K}, curves of growth from
  \protect\cite{2003MNRAS.343.1271J}, A(Li) = 3.3 from
  \protect\cite{1989GeCoA..53..197A}.}
\end{figure*}

These models suggest that lithium depletion can be used as a sensitive
 indicator of age
 in the mass range $0.4 < M_{\star}< 0.6 \,\rm{M}_{\odot}$  ($3400 < T_{\rm{eff}} <
 3800\, \rm{K},\,1.2 < (V-I_{\rm{c}})_{0} < 2.6$). Outside this range, pre-MS stars of
 lower mass
 than $0.4\,\rm{M}_{\odot}$ reach the temperature for lithium burning too
 slowly to show depletion in young $(< 10\, \rm{Myr})$ clusters. Stars of mass
 $>0.6\,\rm{M}_{\odot}$ develop a radiative core early and the base of the
 convective zone rises to regions of cooler temperature before all
 lithium can be destroyed \citep{1997ARA&A..35..557P}, rendering
 lithium depletion ineffective as an age indicator in young clusters.

Studies by \cite{2005ApJ...626L..49P}, \cite{2007ApJ...659L..41P} and \cite{2007A&A...462L..23S}
derive ages based on EW[Li] for individual stars in the ONC and $\sigma$ Ori.
They compare lithium depletion ages for each star with that found using isochrone fitting
in the CMD. A small number of objects are identified in each
cluster with anomalously low lithium line strength and in a few cases these
correlate with an older CMD
age. The lithium-derived age spread found within the clusters is proposed to
support a slow star formation model. However, a small number of stars in the very low mass regime
($\,\lesssim0.25\,\rm{M}_{\odot}$) are found in both the
ONC and $\sigma$ Ori with lithium depletion
significantly greater that expected from their CMD age.

Lithium depletion predicted from simulations of cold
accretion may occur throughout the mass range \cite[see examples in][of
stars with mass 0.1 and $1\,\rm{M}_{\odot}$]{2010A&A...521A..44B} providing the mass accretion rate during bursts
$\dot{M}\ge5\times10^{-4}\,\rm{M}_{\odot}\,\rm{yr}^{-1}$. 
Conversely, age related depletion is highly sensitive to mass
and is shown by some stellar interior models to be suppressed in the
range 0.6 -- 2$\,\rm{M}_{\odot}$ due to the onset of the radiative core.
As a result, the region in EW[Li] vs. colour space between $0.5 > (V-I_{\rm{c}})_{0} > 1.6$
and below a EW[Li] of $0.35\,\rm{\AA}$ is a `forbidden
zone' (shown in Figure 3) where low lithium objects should not exist
according to non-accreting models. Confirmed young cluster members found in the
forbidden zone would offer strong support to cold accretion theories,
even if combined with modest age spreads.

\begin{table*}
 \centering
 %\begin{minipage}

  \caption{Key parameters for clusters selected in this study.}
  \begin{tabular}{@{}lcccl@{}}
  \hline
      Cluster & Age  & Distance  & Number of  & Supporting \\
     &  (Myr) &  modulus & selected targets &  catalogues\\
 \hline
 Orion Nebula Cluster & 2$^{(1)}$ & 8.08$\pm{0.04}^{(2)}$ & 115 &  \cite{1997AJ....113.1733H}\\
 NGC 2264 & 3$^{(1)}$& 9.37$^{+0.15}_{-0.11}  \,^{(1)}$  & 88 & \cite{2007MNRAS.375.1220M} \\

\hline
%\end{minipage}

$^{1}$\cite{2008MNRAS.386..261M}\\
$^{2}$\cite{2007AA...474..515M}\\
\vspace{0.2cm}
\end{tabular}

\end{table*}

\subsection{Target selection}
The constraints established in Section 2.1 lead us to require nearby young clusters with
ages $<10\, \rm{Myr}$, where lithium
depletion may identify pre-MS members with a high accretion rate history or
greater age. We also
required comprehensive literature photometry and spectroscopy to characterise 
the observed stars and allow construction of CMDs for isochronal age
determination. The clusters selected were the ONC and NGC 2264. Key
parameters for these clusters are shown in  
Table 1. Whilst previous age estimates for these clusters are around
2-3 Myr, recent work by \cite{2009MNRAS.399..432N} and \cite{2013MNRAS.428.3178B} suggest that pre-MS ages are underestimated
by a factor 2, hence our clusters may in reality be closer to 4-6
Myr.  

To ensure a high fraction of stars studied
were cluster members, indicators of youth such as X-ray luminosity \citep{1999A&A...345..521F,2002ASPC..277..155F},
H$\alpha$ emission \citep{2005AJ....129..829D}
and optical variability
\citep{2002A&A...396..513H,2004A&A...417..557L} were used in
conjunction with position in colour-magnitude space to
select target stars. Importantly, photometric selection did not
exclude members with low luminosity. Instrumental constraints meant
that selection was also weighted toward
the cluster centre, thus preferring a smaller number of objects close
to the cluster core over a larger number further out. This led to the
selection of fewer stars per field, although those selected had a
higher probability of cluster membership.
 The final selection criteria for target stars was a cut in
 $T_{\rm{eff}}$ \citep[determined via the colour - $T_{\rm{eff}}$ relation
 of][]{1995ApJS..101..117K} in the range
$3400\--4900\,\rm{K}$, approximately corresponding to mass range $0.3 \--1.9\,\rm{
M}_{\odot}$ on the isochrones of \cite{2000A&A...358..593S} at an age
of 3 Myr.

For the ONC, the selection deliberately targeted 7
literature cluster members that appear to be fainter than
the main population in the CMD. Since low-luminosity objects could result
from cold accretion, we shall investigate whether these objects also
exhibit depleted lithium. Alternative theories proposed to explain
them include an
age spread \citep{2000ApJ...540..255P}, as gravitationally captured
field stars \citep{2007MNRAS.375..855P} or unusually blue objects exhibiting
heavy accretion veiling 
\citep{1997AJ....113.1733H}. We shall refer to this population in
the rest of the paper as the `low luminosity group' (LLG).

\section{Observations and data reduction}

The observations were made on the nights of 2011 November 20, 21
using the AF2/WYFFOS multi-object fibre-fed spectrograph on the $4.2\,\rm{m}$ \textit{William
Herschel} Telescope. The small fibres module contains 150 fibres, each
with a diameter of
1.6 arcsec. To avoid potentially damaging fibre collisions, not all of the fibres may be placed
on target stars in any given configuration, thus limiting the actual
number of targets to significantly less than 150 per field. This was an
important limiting factor in these observations due to the centrally
condensed nature of the clusters studied. 
We used an echelle
grating and order sorting filter, centered at $6545\,$\AA\  with a
range of $\sim390\,\rm\AA$. This covered the
wavelength range including H$\alpha$ and the lithium doublet at
$6708\,$\AA. The signal-to-noise (SNR) obtained ranged from 10-60 per 0.22\AA\ wavelength step
with a resolution, R $\sim 10\,000$.

Fibre setups were configured to
maximise the number of cluster members observed. Time, 
weather and observability constraints meant that for the ONC  two
`bright' (1 x $2700\, \rm{sec}$)  and one `faint' (4 x $2700\, \rm{sec}$) setups
were obtained and for NGC 2264 three `bright' (1 x $2700\, \rm{sec}$)
setups. The longer exposure times for the ONC were offset by the
greater levels of extinction to that cluster
\citep[$A_{\rm{v}}\sim1.5-6$, ][]{1997AJ....113.1733H} as compared with NGC
2264 \citep[$A_{\rm{v}}=0.37$, ][]{2007MNRAS.375.1220M}. As a result, the
limiting magnitude (corrected for extinction) for both clusters was
$V_{0} \ge 17$. 115 stars were observed in
the ONC and 88 in NGC 2264.

Science frames were bias
corrected using a median of 5 bias frames, with bad pixels identified
from dark frames and subsequently masked. Tungsten lamp flat-fields were obtained
before each science exposure and used to trace the positions of
spectra on the CCDs. Flat-field images were also used to construct a
weighting function for the optimal extraction routine. Arc lamp
exposures before and after each science frame were used to
provide wavelength calibration. Sky signal was subtracted using a minimum of 30 randomly placed sky fibres.
The spectra were extracted and calibrated using an IDL based software
pipeline provided by the \textit{Issac Newton} Group and
optimised for the AF2/WYFFOS instrument\footnote[1]{\url{http://www.ing.iac.es/astronomy/instruments/af2/reduction.html}}. Spectra were extracted using an optimal
routine \citep[see][]{1986PASP...98..609H} which helps to minimise
statistical noise. Once calibration and reduction were complete the data were then
interpolated and re-binned onto $10\,000$ logarithmically placed
wavelength steps.

Target spectra, measured relative radial velocities, lithium equivalent
widths and veiling fractions in this paper are freely available from
the Cluster Collaboration home
page\footnote[2]{\url{http://www.astro.ex.ac.uk/people/timn/Catalogues/}}
and the CDS archive (a sample may be seen in Table 2).

\section{Analysis}

The $6708\,$\AA\ lithium line equivalent width (EW[Li])  was measured with
respect to a pseudo-continuum. A wavelength step integration of the
spectral line was performed between upper and lower line limits
defined to exclude other lines and to incorporate the entire line
irrespective of rotational broadening or binarity. At the resolution
of our spectra, contamination of up to $20\,\rm{m\AA}$ is likely to be
present from a
weak Fe I line at 6707.4\,\AA. Uncertainty on definition of the
pseudo-continuum and line limits were determined through
repeat measurements to be $\pm\,10\,\rm{m\AA}$.
Photon noise in the signal contributes a further
$\pm\,10-60\,\rm{m\AA}$, dependent on the SNR of the spectrum.  The
three uncertainties are combined in quadrature with that described in
Section 4.2 for accretion veiling.

As we are trying to measure the intrinsic EW[Li] for these stars, we
must account for `veiling' of spectral lines by continuum emission from magnetospheric accretion regions on the surface of the star. This
`veiling' effect is well studied
\citep[eg.][]{2008A&A...482L..35G,2012arXiv1209.1851D} and its effect has been corrected through the
use of techniques outlined in \cite{1989ApJS...70..899H}. The
accretion veiling contamination is visible as a weakening of
absorption lines and hence has a significant effect on the measurement of
photospheric features such as the 6708\,\AA\ lithium line.
The effect of accretion veiling is to reduce measured EWs, hence
without careful consideration this could affect the apparent lithium
abundance in the objects being studied. A correction for the veiling
flux is made assuming that it is a smoothly varying continuum over the 
spectral range from 6390 to 6710\AA. Within the narrow range
specified this is a reasonable assumption as determined by
\cite{1989ApJS...70..899H} and \cite{1994ASPC...62..132G}.
Before determining and correcting for veiling, an assessment of radial velocity
and $v\,$sin$\,i$ must first be made between the object and template
stars in order to correct for wavelength and line profile
differences. Typically the unveiling process contributes an uncertainty of similar
magnitude to the combined measurement uncertainties and is combined
with them to provide an uncertainty estimate on the final measurements.

\subsection{Radial velocity and binary stars}

Radial velocities (RVs) were determined using
cross-correlation of spectral lines in the range $6385\--6510\,\rm{\AA}$ between each star and an arbitrary
(low $v\,$sin$\,i$) single star reference in the field. No absolute RV standards were measured
as relative velocities are sufficient for veiling analysis and cluster membership
verification.  Figure 4 shows histograms of the derived radial
velocity distributions for both clusters.

\begin{figure*}
\begin{center}  
\includegraphics[width=\columnwidth]{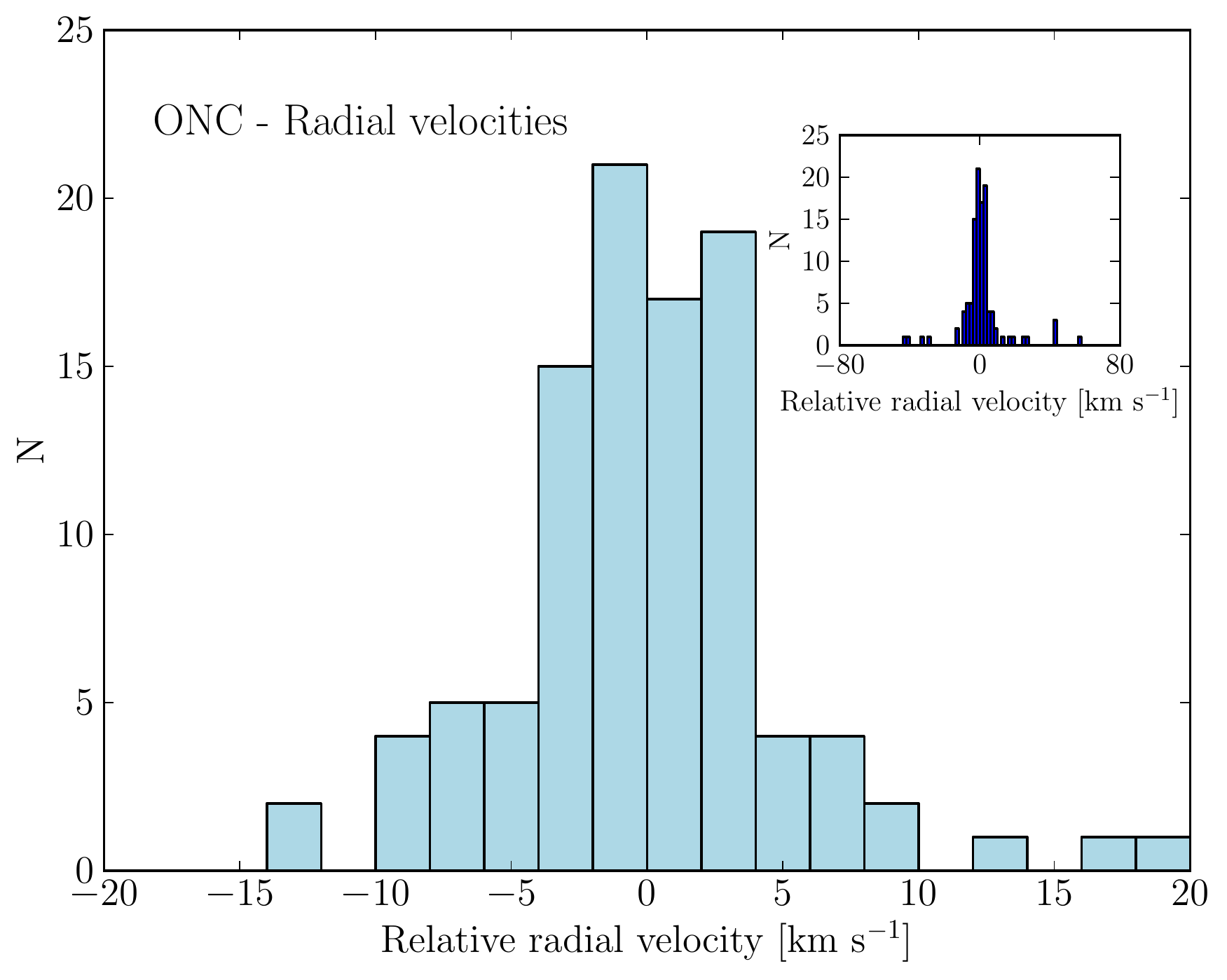}
\hspace{0.5cm}
\includegraphics[width=\columnwidth]{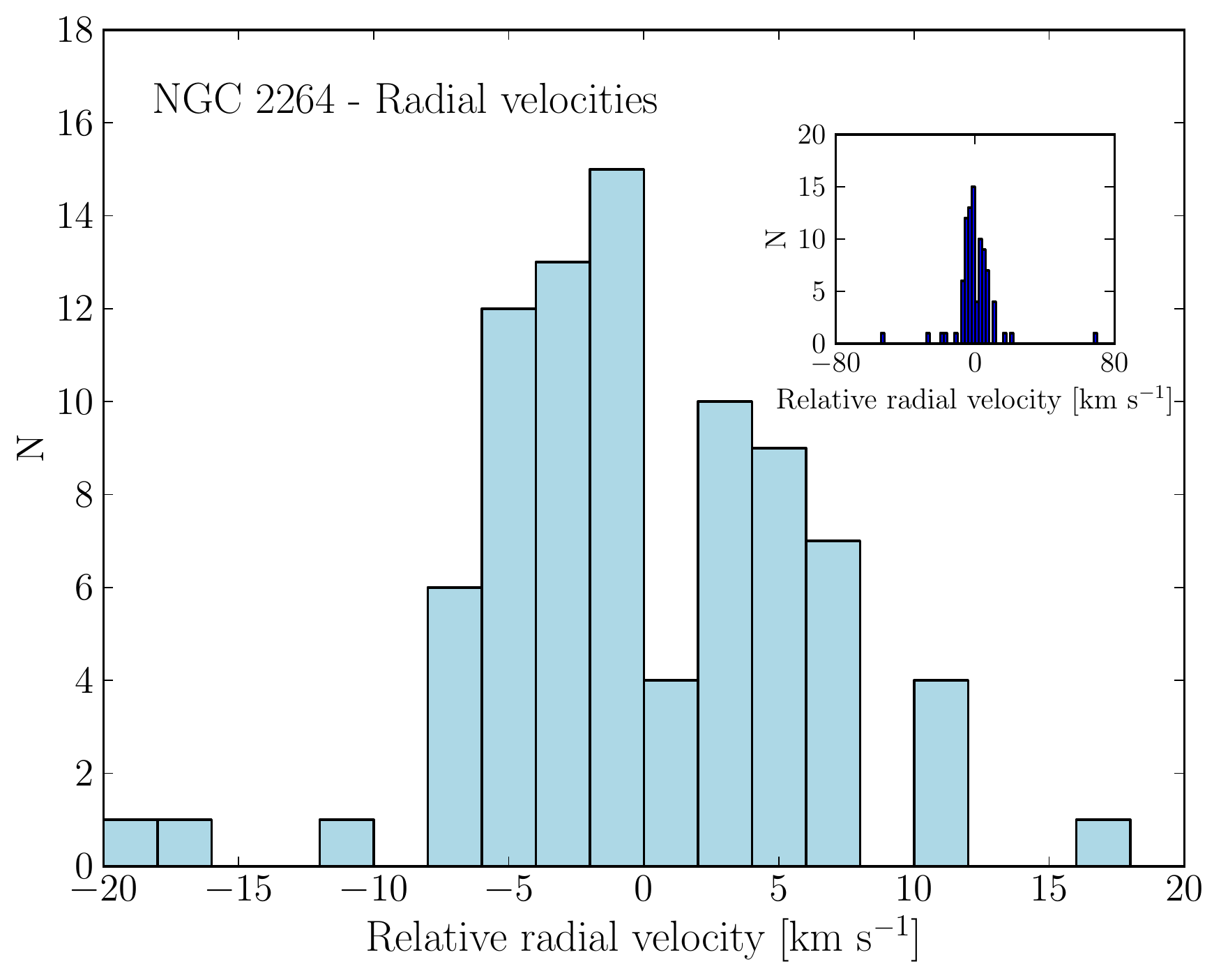}
\end{center}
\caption{Histogram of radial velocities for stars in the ONC (left) and
  NGC 2264 (right) relative to the median of the respective samples.
Inset shows the entire sample including radial velocity outliers.}
\end{figure*}

The ONC displays a clear velocity grouping with a range of
$\pm\,10\,\rm{km}\,\rm{s}^{-1}$ about the sample median, this is consistent with previous RV studies of
this cluster \cite[e.g.][]{2009A&A...508.1301B}.
21 objects were identified with velocities outside the range specified and hence were removed from
further analysis of the sample. Many of these objects are likely to be unequal mass spectroscopic binary cluster members that are unresolved from their
cross-correlation function (CCF), however without multi-epoch measurements
we cannot be clear as to which objects fall into this category. As a
result we reject all stars with $|\rm{RV}| >\, 10\, \rm{km}\, \rm{s}^{-1}$.

The NGC 2264 RV histogram shows that this cluster exhibits a
non-Gaussian velocity dispersion with a range of $\pm\,9\,\rm{km}\,\rm{s}^{-1}$ about the median. This is consistent with
previous measurements by \cite{2006ApJ...648.1090F}. 9 objects in the
NGC 2264 sample exhibit RV with magnitudes outside
the range defined above and hence have also been removed from further
analysis.

Some of the identified RV outliers exhibit lithium absorption and
H$\alpha$ emission, so are likely to actually be cluster members. We have
been deliberately cautious in excluding potential non-members though, as our
aim was to set up a robust sample of members that does not use
lithium detection as a criterion. Exclusion of some stars which are
actually members only affects the result by reducing the sample size. Conversely, including a star that was not a member could seriously bias the result.

Binary stars with resolved splitting in spectral lines cause significant difficulties with accretion spectrum
fitting. Potential binaries were identified since they exhibit
multiple peaks in the CCF and were removed from the data set before
the veiling
analysis. Examples showing
single and binary star CCFs are shown in Figure 5.
Multiple peaks in a broad CCF could also be due to noise overlaid on
highly rotationally broadened spectra. This is difficult to confirm or refute on a
star by star basis, however unveiled Li[EW] measurements for these
stars were included in the analysis as veiled lower limits in EW[Li].

\begin{figure*}
%\begin{center}  
\includegraphics[width=5.5cm]{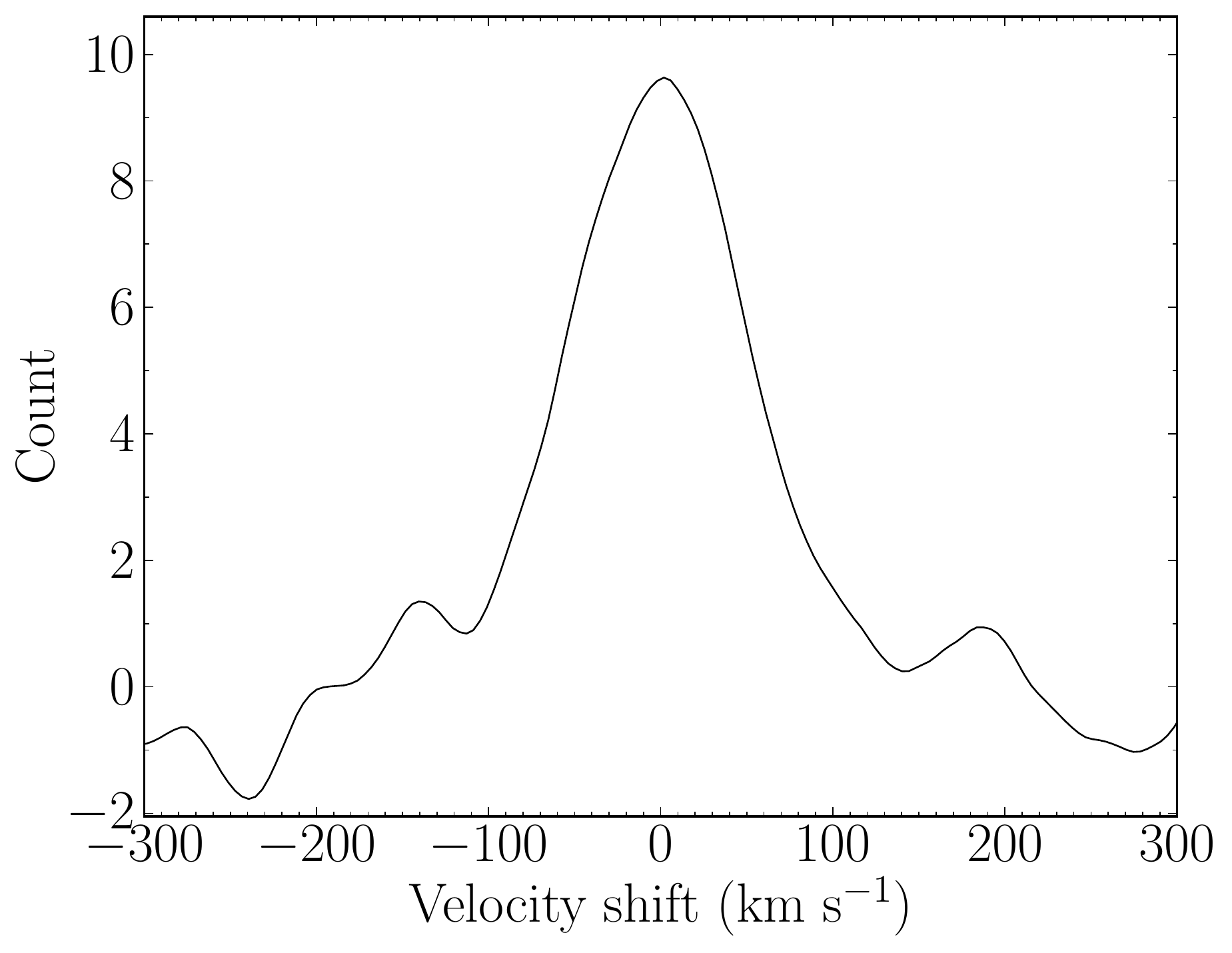}
\includegraphics[width=5.5cm]{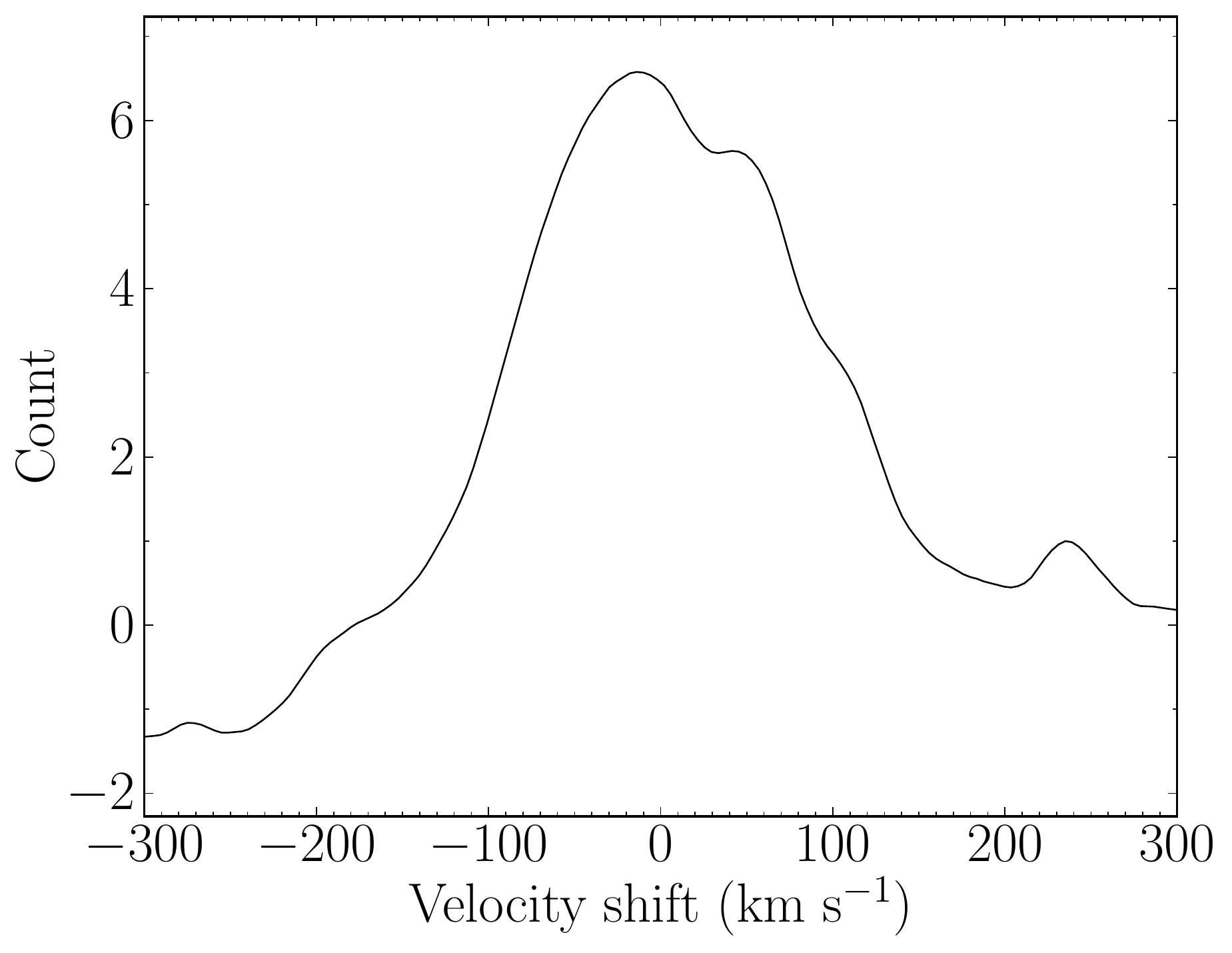}
\includegraphics[width=5.5cm]{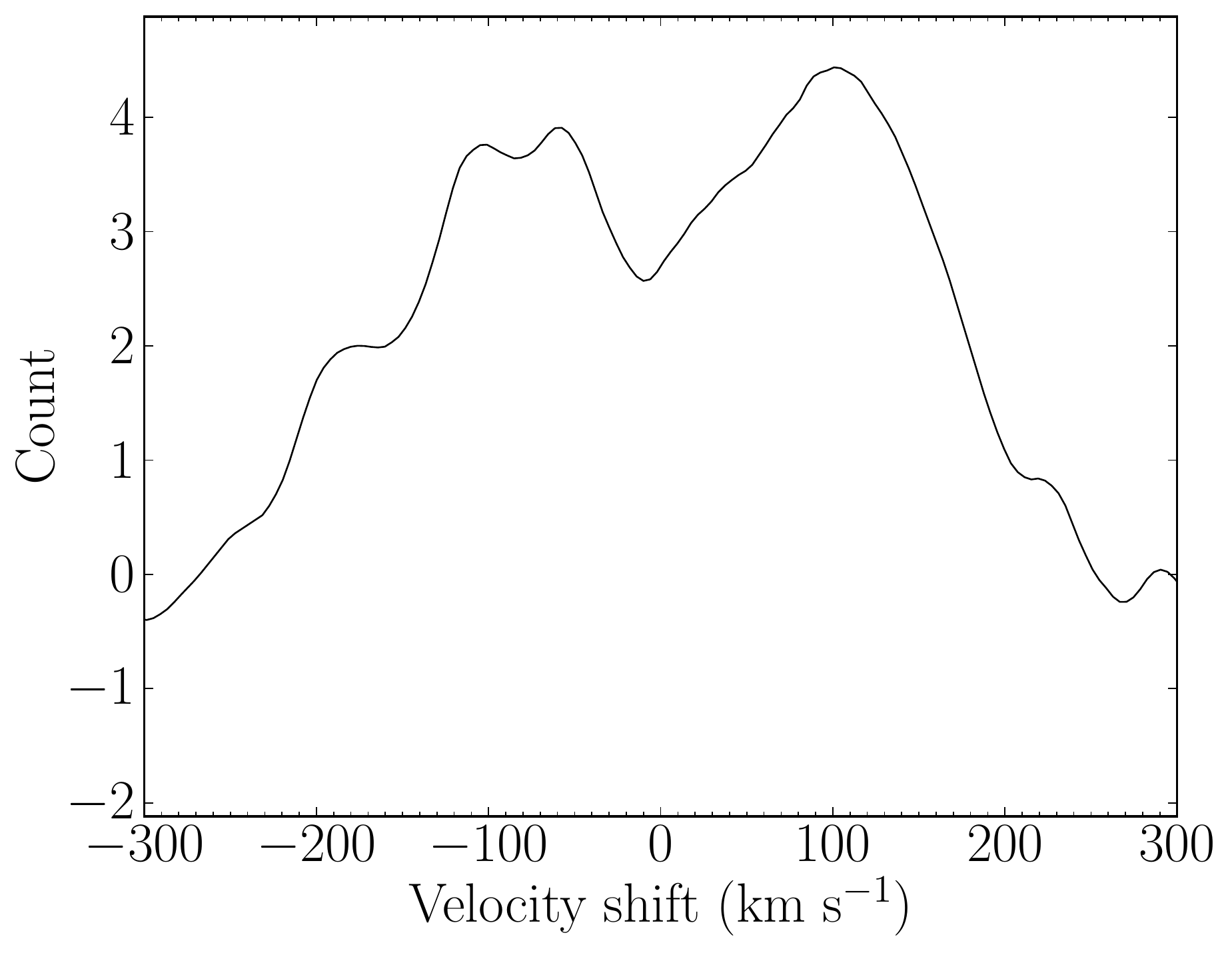}
%\end{center}
\caption{Example velocity cross-correlation function (CCF) plots, showing single star
  (left), and close and wide line separation binary stars (centre and right).}
\end{figure*}

Four of the seven stars in the ONC `LLG' (see Section
2.2) were
identified as having RVs that are inconsistent with the
cluster. As a result they are removed from the sample and their
faint nature may be explained as being characteristic of main-sequence
or post-main-sequence
stars from the galactic field that are moving through or behind the
cluster. These objects are illustrated by red crosses in Figure 1.

\subsection{Accretion veiling and $v\,$sin$\,i$}

 The magnitude of veiling for a given object
over wavelength range $j$ is expressed as:
\vskip 0.3cm
\begin{equation}
r_{j} = \frac{k_{j}} {S_{\rm{phot}}} 
\end{equation}
\vskip 0.3cm
\noindent where $r_{j}$ is the flux ratio of the accretion veiling continuum ($k_{j}$) to the stellar photospheric continuum ($S_{\rm{phot}}$).

To estimate the veiling continuum flux,
a set of template stars have been selected for each cluster from their
own populations (9 for the ONC
and 8 for NGC 2264) across the full colour range. Selected template
stars form the upper envelope of the EW[Li] vs. $(V-I_{\rm{c}})_{0}$ diagram and exhibit $v\,$sin$\,i$
limited by instrument FWHM. Hence these objects should
be amongst those with the lowest veiling and intrinsic $v\,$sin$\,i$ within our sample.
The templates were first offset in
wavelength to compensate for the difference in RV between
template and object. They were then rotationally broadened
\citep[using a kernel with limb darkening coefficient of
0.5;][]{1993AJ....106.2096V}
and normalised to a similar flux as that of the object spectrum to
produce reference templates.

The templates were then subtracted from the
object spectrum to leave residuals which are assumed to represent the
accretion spectrum. We varied the normalisation of the templates
and examined the residuals using a $\chi^{2}$ test, performed between
the residuals and ($30\,\rm{\AA}$ boxcar average) smoothed versions of themselves.
Adopted $v\,$sin$\,i$ and veiling values were identified using the residuals with
minimum $\chi^{2}$. The parameter search space was defined by $v\,$sin$\,i$, from $0\--\,100\,\rm{km}\,\rm{s}^{-1}$ with a
resolution of $1\,\rm{km}\,\rm{s}^{-1}$ and veiling ratio of 0 -- 5 in steps of
0.01 of the photospheric continuum. Figure 6  shows an example of an
object spectrum, template and best fitting broadened and veiled model
template.

This process was repeated as it was found that a considerably
stronger CCF
was obtained if the RV cross correlation is performed
against an appropriately broadened template. Thus the first iteration
identifies approximate RV and $v\,$sin$\,i$, the second iteration
improves the RV and hence $v\,$sin$\,i$ as the fitted RV template has
a more similar shaped line profile.

Fitting of templates was performed in two wavelength regions, the first between 6390 and 6500\AA, in a region containing 7 photospheric absorption lines (Fe I  6393.6\AA, Ne II
6407.9\AA, Fe I  6411.7\AA, Fe I  6421.3\AA, Fe I  6431.0\AA, Ca I
6439.1\AA, Ca I
6450.0\AA, Ca I 6462.5\AA\ and Fe I 6495\AA). The second was between
6610 and 6675\AA, containing three lines (V I 6624.8\AA, Ni I
6643.6\AA\ and Fe I 6663.4\AA). These regions are free from
contamination by H$\alpha$ or significant sky and nebular line emission.

Analysis of sky and nebular lines indicate that the
spectrograph was operating with a resolution of R
$\sim10\,000$. Comparison with literature \citep[see][]{2001AJ....122.3258R,2009AJ....138..963B}
indicate that our limit in $v\,$sin$\,i$ resolution is
$\sim20-30\,\rm{km}\,\rm{s}^{-1}$, depending on target SNR.

The veiling for each star was calculated as the
weighted mean across the cluster template set based on the inverse of the
difference in $(V-I_{\rm{c}})_{0}$ colour between each template star
and the object. To avoid overweighting when the template and object
colours were very close, a minimum $\Delta (V-I_{\rm{c}}) _{0}$ of 0.1 was placed on
the weighting algorithm. The aim of this process is to use multiple
templates to provide reliability of fit, without introducing errors
from large template-to-star colour differences. It should be noted
that despite the efforts taken to ensure that the veiling was derived
from templates of similar colour, in practice the veiling measured
was found to be surprisingly insensitive to template colour for any
given star. 
The range of fractional uncertainty due to template mismatch was measured to be $\sim 0.05 - 0.3$ across our sample.
Template to template variations appear to dominate, possibly caused by
differences in the template line profiles due to rotation, binarity, and chromospheric activity.

Uncertainties in veiling were derived by analysing the range
in veiling estimates returned by all of the fits from the full template set. The standard deviation (also weighted as a function
of template to star colour difference) of the set of veiling values was calculated and adopted as the
uncertainty in the veiling measurement. 

As the method used here analyses lines over a narrow spectral range, it
is independent of reddening and extinction and hence is robust in
dusty regions of new star formation such as the clusters studied in
this paper. Measured
values of $r_{j}$ vary from 0 to 4.0.

\begin{figure}
%\begin{center} 
\includegraphics[width=\columnwidth]{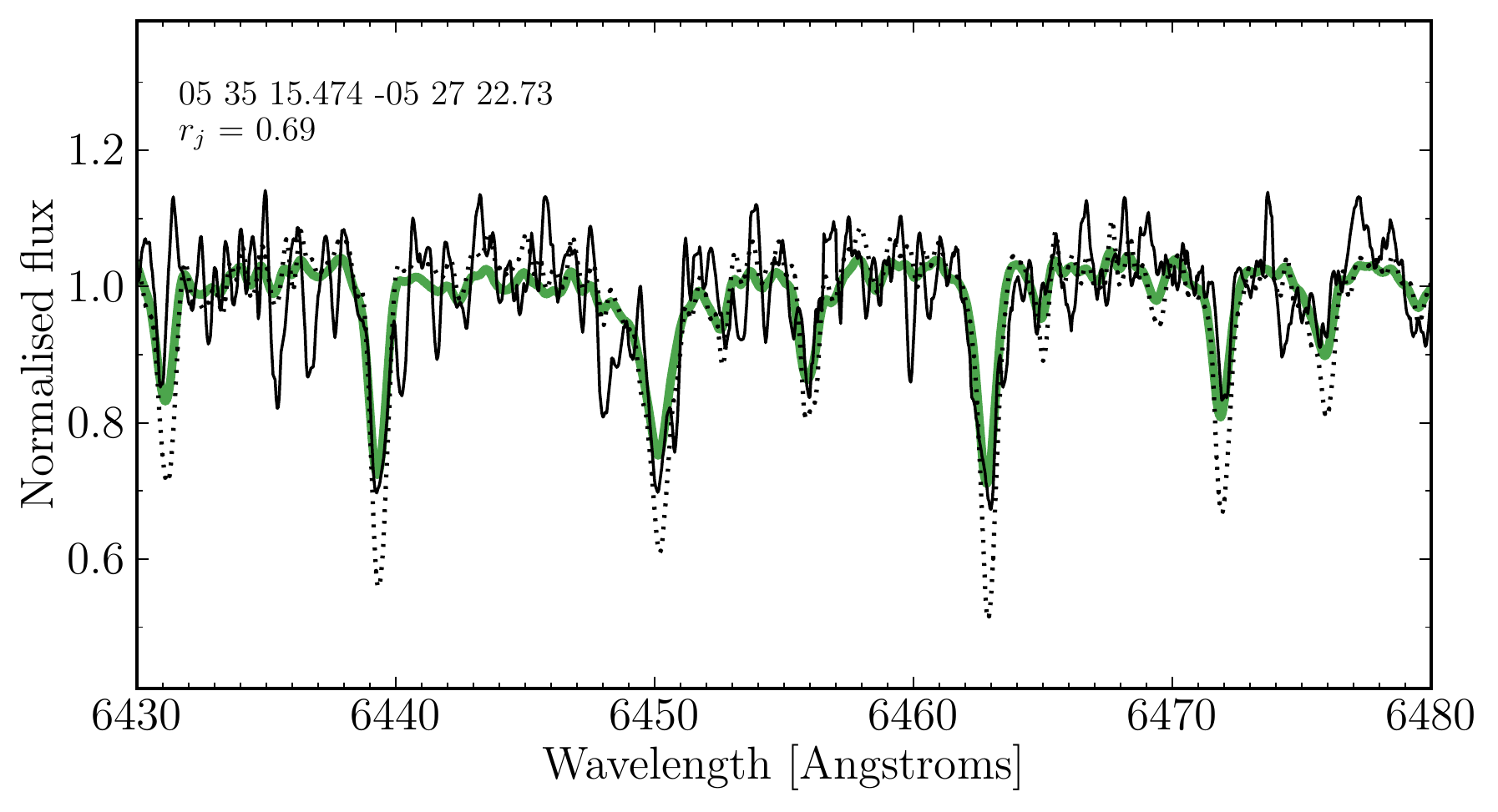}
%\end{center}
\caption{Spectrum of object RA = 05 35 15.47 Dec = -05 27 22.7 in the ONC. The thin
  black line is the object spectrum. The
  dotted line is the template showing significantly deeper
  photospheric absorption lines. The green (light grey) broad line is
  the broadened and veiled template that best fits the object
  spectrum ($r_{j}$ = 0.69).}
\end{figure} 

\subsection{Lithium equivalent width vs. colour}
Once the magnitude of the veiling continuum had been established for each star,
a correction was applied to the raw measured lithium EW to quantify the
unveiled EW. The uncertainty in the corrected EW[Li]
was calculated by combining uncertainties in the raw line measurements with
the uncertainty in veiling. Figure 7
shows veiling corrected lithium EWs and associated uncertainties plotted as a function
of dereddened $(V-I_{\rm{c}})_{0}$ colour
{\rm Most of our data points lie slightly
above the undepleted model isochrone. Reasons for this may include
uncertainties in molecular line strengths in stars with
$T_{\rm eff} < 4000K$, causing difficulty in definition of the the pseudo-continuum around the lithium line. 
In the hotter stars a lack of inclusion of NLTE effects may cause
uncertainties in the modelled curves of growth \citep[e.g.][]{1994A&A...288..860C}.
}Theoretical 0 (undepleted), 10 and 20 Myr isochrones
\citep{2002A&A...382..563B} are also shown. Dereddened $(V-I_{\rm{c}})_{0}$ colours for the ONC are from
\cite{1997AJ....113.1733H}. It should be noted that the dereddening
method used for the ONC was performed on a star-by-star basis and groups objects into bins by spectral class. Hence
$(V-I_{\rm{c}})_{0}$ colours are not continuous. Dereddened colours for
NGC 2264 were derived in \cite{2007MNRAS.375.1220M} using a single
global solution of $E(V-I_{\rm{c}})$ = 0.15, based on a mean measured $(V-I_{\rm{c}})$ 
excess for main-sequence stars.

Accurate unveiling of spectroscopic binaries is a very
difficult and uncertain process, so these points have been left in
their raw measured (veiled) values and are displayed as lower limits
in EW[Li].

\begin{figure*}
\begin{center}  
\includegraphics[width=\columnwidth]{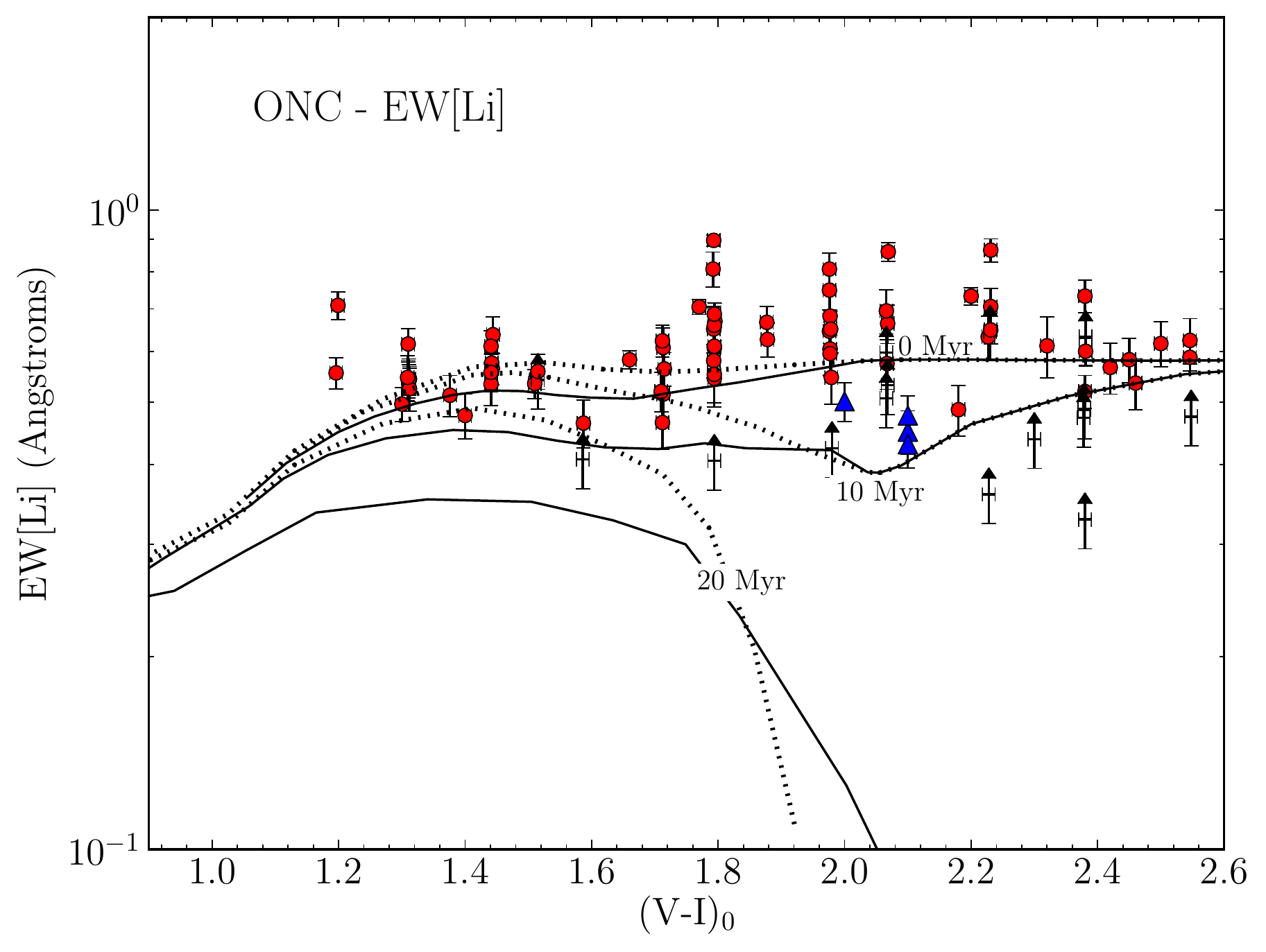}
\hspace{0.5cm}
\includegraphics[width=\columnwidth]{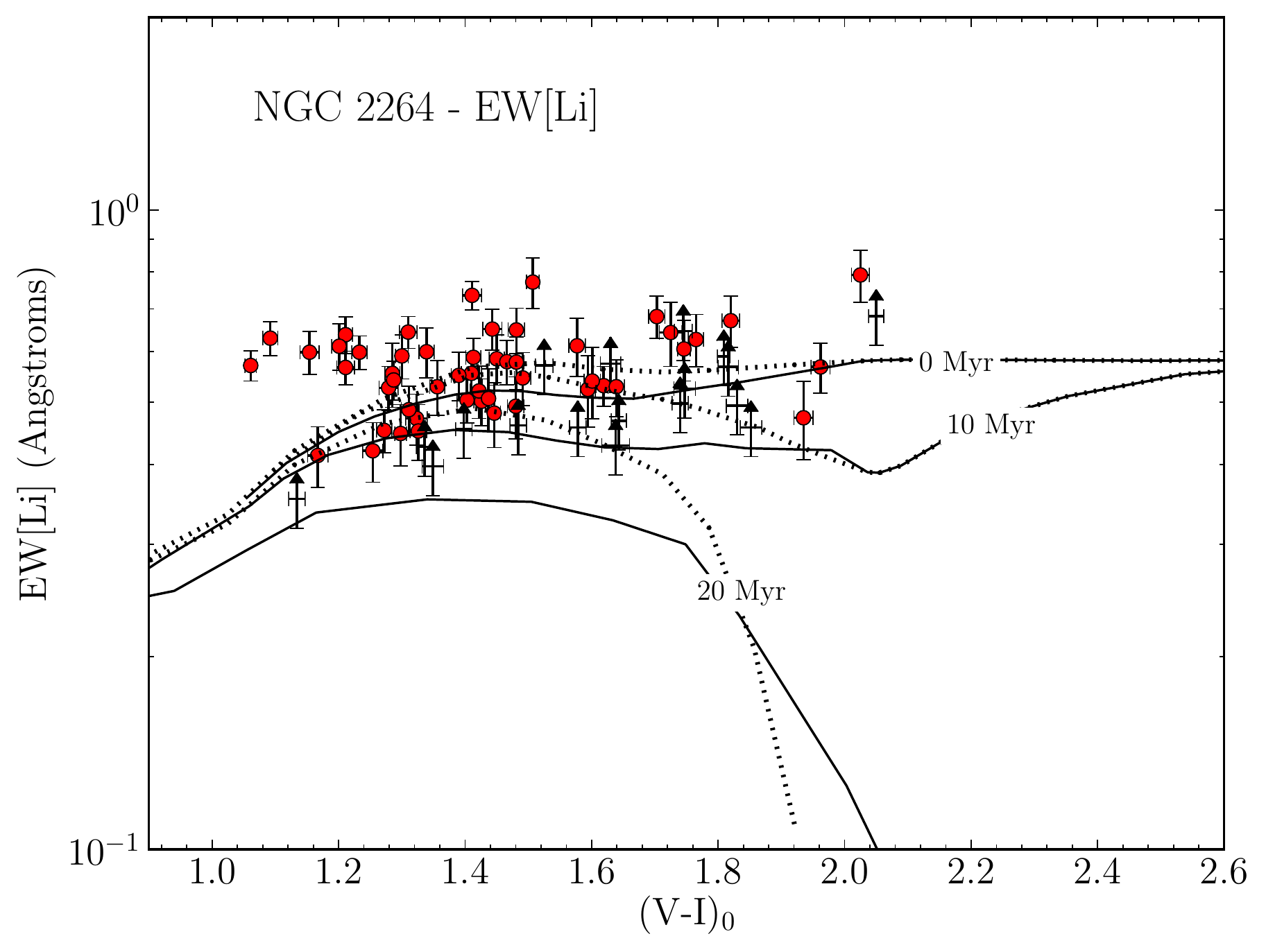}
\end{center}
\caption{A comparison of $6708\,$\AA\ lithium line equivalent width
  measurements for stars in the ONC (left) and NGC 2264 (right)
  plotted as a function of intrinsic $(V-I_{\rm{c}})_{0}$. Red points are
  unveiled single star measurements, black triangles are lower limits
  for spectroscopic binaries which are left as veiled (raw)
  measurements. The 0 Myr, 10 Myr and 20 Myr isochrones are based
    on models by \protect\cite{2002A&A...382..563B}. Solid lines indicate models using a convective mixing length parameter ($\alpha$) of 1.9. Dotted lines indicate $\alpha$ = 1.0.  Blue triangles are data from
  \protect\cite{2005ApJ...626L..49P} which are cited as having
  depleted lithium. These points have been calculated using n(Li) to
  EW[Li] conversion from curves of growth in Table 1 of
  \protect\cite{2007ApJ...659L..41P}. The conversion from
  $T_{\rm{eff}}$ to intrinsic
  $(V-I_{\rm{c}})_{0}$ colour for the
  \protect\cite{2005ApJ...626L..49P} data have been made using the
  relation defined by \protect\cite{1995ApJS..101..117K}.}
\end{figure*}

{\subsection{Hydrogen $\alpha$ measurements}

The spectral range of our data included the H$\alpha$ line at
$6563\,\rm\AA$. Whilst it is beyond the scope of this paper to analyse this
feature, measurements of H$\alpha$ equivalent width and the velocity width
at 10\% of peak line flux have been measured and are included in the tables
of data for future reference. Importantly, the H$\alpha$ equivalent width
measurements presented are not corrected for accretion veiling. Sky
and nebular line subtraction for this data set has been performed
using a median combination of randomly placed sky fibres and at the
resolution of our study, structure within the H$\alpha$ line is poorly
resolved. These two factors mean that nebular line contamination of
H$\alpha$ measurements is likely to be significant \citep[see][]{2008ApJ...676.1109F}
and thus these data should be
used with caution.

Data from this paper are available online as an electronic table. An
example of the data format is shown in Table 2. Spectra are also
available online for all of our targets. 

\begin{table*}
\begin{center}
\caption{Summary of derived parameters from this study. Table 2 is
  published in its entirety in the electronic edition. A portion is
  shown here for guidance regarding its form and content. Coordinates
  are (J2000.0). Importantly, the resolution and sky
  subtraction scheme used in this analysis mean that nebular line contamination of
H$\alpha$ measurements is likely to be significant and thus H$\alpha$ data should be
used with caution. EW[H$\alpha$] measurements are not
  corrected for veiling. }
\begin{tabular}{cccccccccccc}
\hline

RA&Dec&EW[Li]&$\Delta$EW[Li]&RV&$\Delta$RV
&$r_{j}$&$\Delta\,r_{j}$&EW[H$\alpha$]&$\Delta$EW[H$\alpha$]&vel
H$\alpha$&$\Delta$vel
H$\alpha$\\
hh mm ss.ss&dd mm ss.ss&\AA&\AA&km\,s${^{-1}}$&km\,s${^{-1}}$&-&-&\AA&\AA&km\,s${^{-1}}$&km\,s${^{-1}}$\\
\hline
05 35 16.32&-05 15 38.00&0.667&0.039&0.5&2.8&0.29&0.04&-45.50&0.03&71.76&1.83\\
05 35 04.20&-05 15 21.44&0.596&0.046&4.8&2.8&0.59&0.10&-142.01&0.08&80.90&1.83\\
05 34 52.25&-05 12 03.20&0.625&0.051&-3.8&2.8&0.14&0.04&-32.50&0.05&82.73&1.83\\
\hline

\end{tabular}
\end{center}
\end{table*}

\section{Discussion}

\subsection{Accretion veiling}

Our data show that our sample of stars in the ONC exhibit
greater accretion veiling than our sample in NGC 2264.
The mean r$_{j}$ and standard deviation have been calculated for all
stars in both clusters and for a subset in both clusters that cover a common colour range ($1.2 < (V - I_{c})_{\rm o} < 2.0$).
All measured stars in the ONC exhibit a mean r$_{j}$ = 0.48 with a
standard deviation ($\sigma$) of 0.68. This compares
with mean r$_{j}$ = 0.16, $\sigma$ = 0.19 for all measured stars in
NGC 2264.
The subset with colour range $1.2 < (V - I)_{\rm c} < 2.0$ in the ONC
exhibit a mean r$_{j}$ = 0.44, $\sigma$ = 0.71 and in NGC 2264 the mean r$_{j}$ = 0.15, $\sigma$ = 0.18.
It is notable
that whilst the absolute levels of veiling differ, the ratio of $\sigma
/ r_{j}$ is similar in both clusters.

If NGC 2264 is older than the ONC as suggested through studies of young main-sequence stars \citep{2008MNRAS.386..261M} then our finding is
consistent with models of decreasing disc fraction (and hence
accretion) with age. If however, the ages of the clusters are
indistinguishable from each other as suggested by main-sequence
turn-off ages \citep[see][]{2009MNRAS.399..432N} then the difference
in apparent accretion luminosity may suggest an environmental
difference between the two clusters such that stars in the ONC 
accrete more vigorously or for longer than NGC 2264.

\subsection{Accretion veiling correlation with CMD position}

To test for any correlation between veiling and position in the CMD, each
cluster was split onto two sub-samples representing the bright and
faint populations using the following method. First we calculated the
median dereddened colour in 0.1 mag. wide magnitude bins, and a straight line was
fitted to define the median magnitude as a function of colour. These
lines may be seen plotted over the data in Figure 1. Stars
with magnitudes brighter than this were assigned to the bright
sub-sample and vice-versa. Histograms of veiling strength in the two
luminosity samples are
shown for each cluster in Figure 8.

We tested the null-hypothesis that the bright and faint sub-samples
were drawn from the same parent veiling distribution using a 2 sample
K-S test. For the ONC, the null hypothesis was rejected with 64 per cent confidence that the
populations are distinct, with a bright median $r_{j}$ = 0.24 and
a faint median $r_{j}$ = 0.23. The implication is that there is
no correlation between veiling and CMD position for the ONC.

The null hypothesis was rejected with 98.3 per cent
confidence for NGC 2264. This suggests that
the brighter stellar population exhibits greater levels of
accretion (median bright $r_{j}$ = 0.17, median faint $r_{j}$ =
0.09). The histogram shown in Figure 8 indicates that this result is mostly
driven by just a few veiled objects in the bright
population.

\begin{figure*}
\begin{center}
\includegraphics[width=\columnwidth]{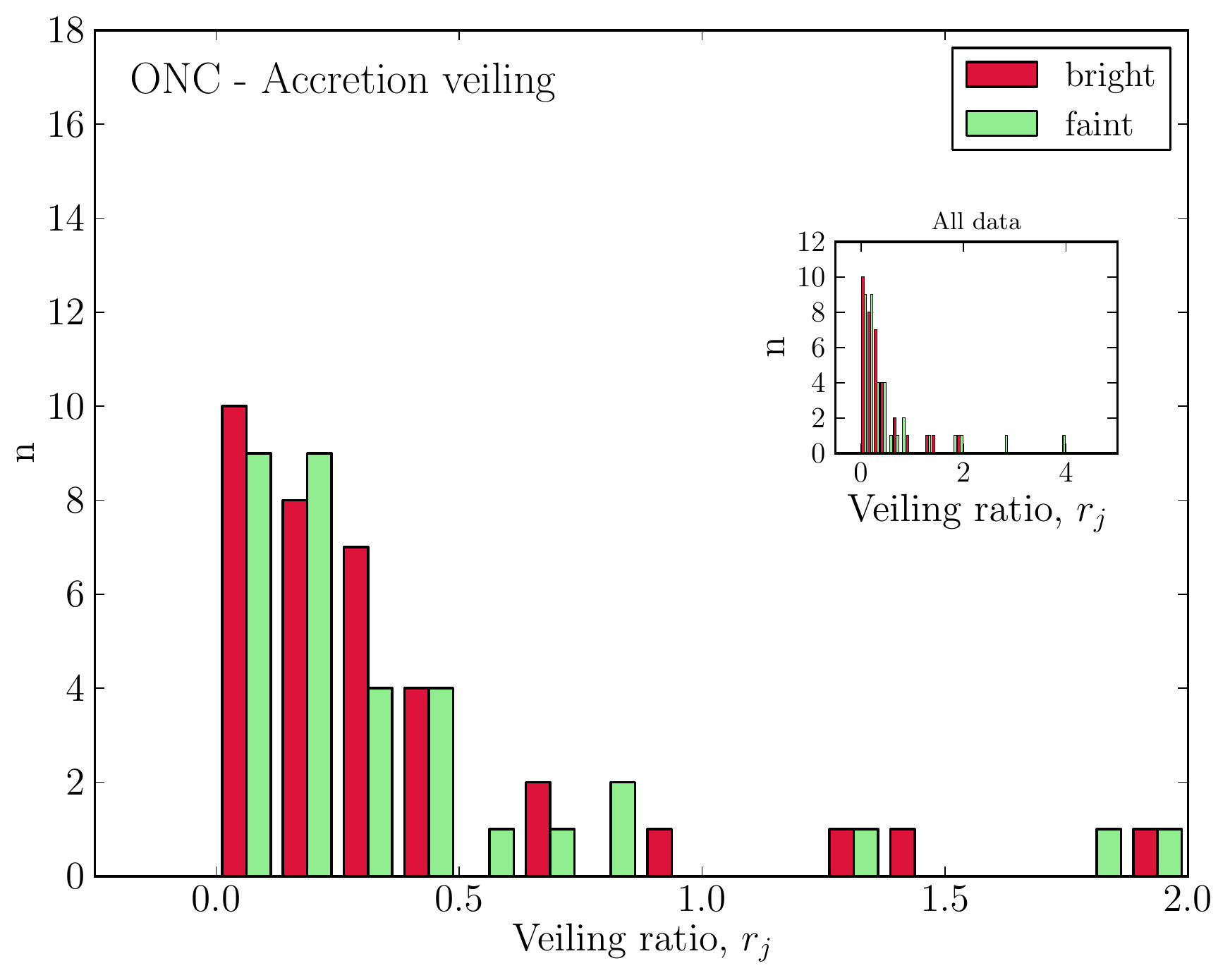}
\hspace{0.5cm}
\includegraphics[width=\columnwidth]{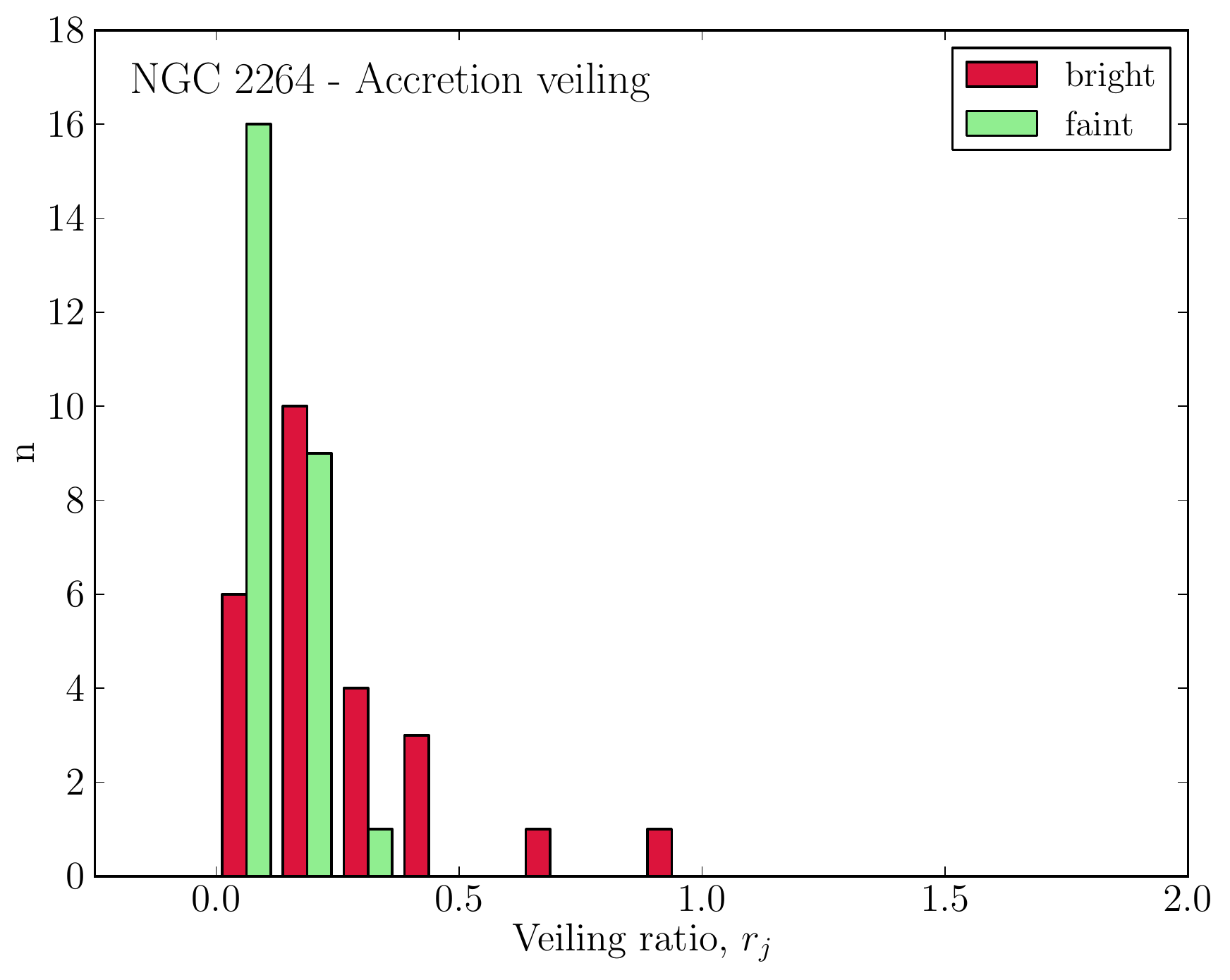}
\end{center}
\caption{Comparison of accretion veiling ratio $r_{j}$ as a function
  of CMD luminosity sample
for the ONC (left) and NGC 2264 (right). Each cluster is split
into a bright and faint population as described in Section 5.2.
Note the axes used for both clusters are the same to aid comparison,
however the ONC exhibits an extended tail of objects at high veiling
ratios (shown in the inset plot).}
\end{figure*}

\subsection{EW[Li] distribution, age spreads and correlation with CMD position}

Both clusters appear to exhibit lithium equivalent widths that are
approximately consistent with their CMD derived ages
($<\,10\,\rm{Myr}$). It is also apparent that no highly lithium
depleted stars are seen with
$\rm{EW[Li]}\lesssim0.4\,\rm{\AA}$. NGC 2264 appears to show a slightly
greater spread in EW[Li] at higher masses than the ONC.

\begin{figure*}
\begin{center}  
\includegraphics[width=\columnwidth]{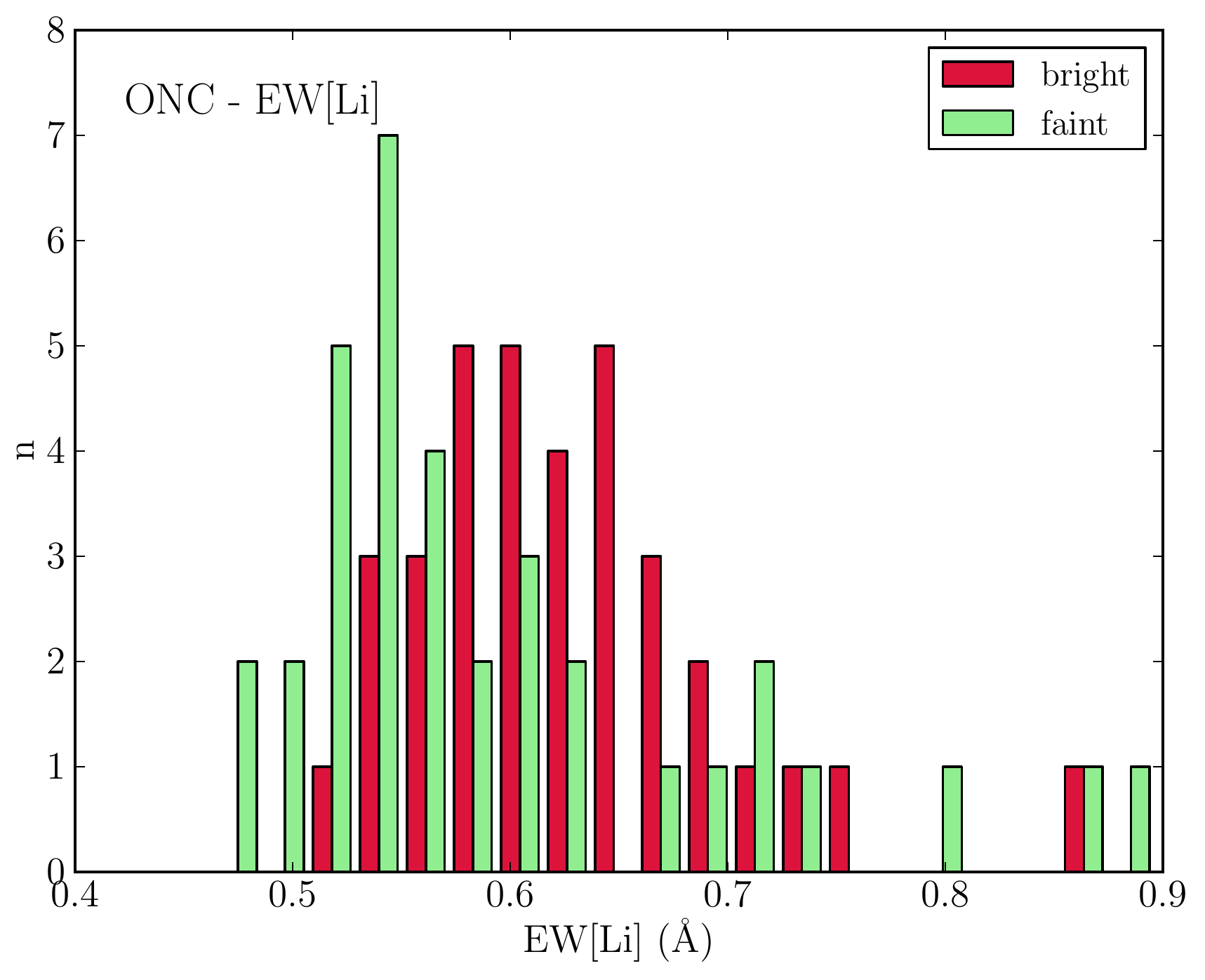}
\hspace{0.5cm}
\includegraphics[width=\columnwidth]{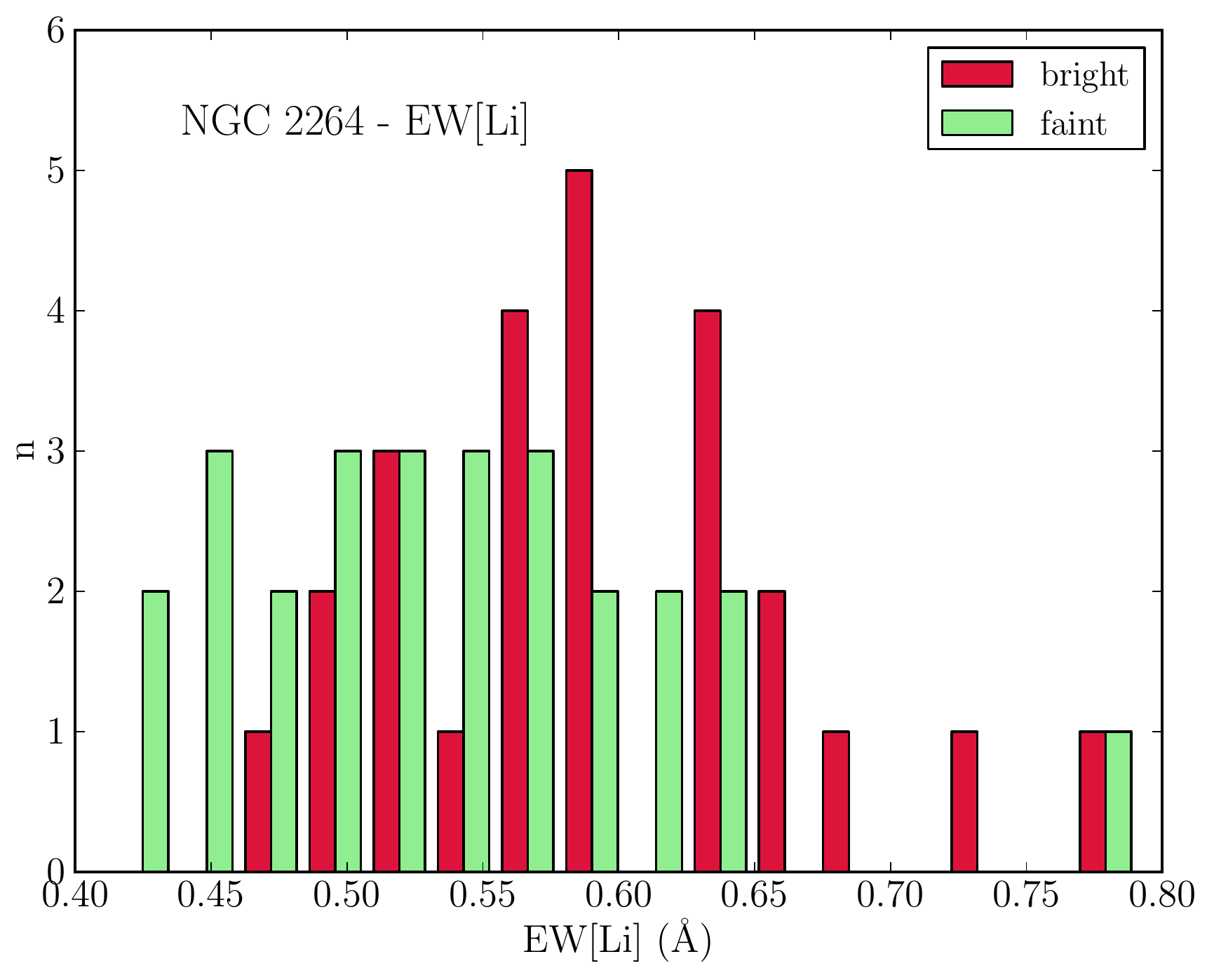}
\end{center}
\caption{Comparison of the EW[Li] data for bright and faint
  sub-samples in each cluster. In both clusters the population of
  faint stars shows a systematic decrease in lithium line strength.}
\end{figure*}

The observed dispersion in EW[Li] might imply an age spread that is similar
in magnitude to the apparent age spread implied by the luminosity dispersion seen
in CMD space. To test whether this really could be an age spread, we looked
for disagreement between the EW[Li] seen in the bright and faint sub-populations defined
previously for accretion veiling (see Section 5.2). If an age spread
explanation were to be plausible, the bright sample should
be the youngest and ages should increase as one moves
to fainter magnitudes. This should be evident as a depletion in
lithium in the faint population.

We compared the sub-populations for each cluster using the 2
sample K-S test. The ONC shows evidence for the two samples
being drawn from different populations with the null hypothesis that
they are from the same population being rejected at the 99.8 per cent confidence level. The bright
population exhibits median EW[Li] = 0.63$\,$\AA\ and the faint
population median EW[Li] = 0.56$\,$\AA.  A similar trend is shown
in NGC 2264, the bright population exhibits median EW[Li] =
0.59$\,$\AA\ and the faint population EW[Li] = 0.54$\,$\AA\ with the
null hypothesis being rejected at the 98.2 per cent confidence
level.
These findings are illustrated in
Figure 9.

We have carefully considered whether this trend in EW[Li] is a
systematic effect of the unveiling process. 
Although there is a weak correlation between EW[Li] and veiling, this cannot be the cause of the correlation of CMD position with Li[EW] since there is not the correlation between veiling and position in the CMD
which would be required to translate this into the observed effect.
It should be noted that if there is a veiling related systematic error, one would expect
its effect to be most apparent in the ONC where veiling is
strongest. In practice the ONC shows no correlation between veiling
and CMD position, yet the EW[Li] vs. CMD position correlation is
significant.

We have also considered whether this trend in EW[Li] may be due to
differences in
surface gravity, however the correlation is in the
opposite sense to that found in simulations by
\cite{2002A&A...384..937Z} and \cite{2005ApJ...626L..49P} as a function of log$\, g$. Another consideration is whether the
observed depletion (either real or a systematic measurement error) is related to rotation rate. Studies such as
\cite{2012arXiv1209.1812V} suggest that rotationally induced mixing can
deplete lithium on timescales of just a few Myr. We observe no
correlation when
EW[Li] is plotted against measured $v\,\rm{sin}\,i$. The depletion we
see would also conflict with the rotation
rate vs. CMD position correlation seen by \cite{2011MNRAS.413L..56L}
if rotationally induced mixing were responsible.

Interpreting this trend in luminosity and EW[Li] is critically
dependent on the adopted value of the convective mixing length
($\alpha$). In the model where $\alpha =
1.9$, the EW[Li]
trend is compatible with an evolution where ageing stars
are slowly depleting lithium across the $(V-I _{\rm{c}})$ colour range as
they contract toward the zero age main-sequence (ZAMS). This model agrees with \cite{2007MNRAS.381.1169J} who finds a statistical
decrease in radius as one moves from bright to faint objects at a
given colour in the CMD.
Importantly, if $\alpha$ is closer to 1.0 then we would expect to see
a difference in dispersion with colour if a large ($\geq 10$ Myr) age spread were present
as at $(V-I _{\rm{c}})
\approx 2.2$ the uncertainties on EW[Li] are less than the
depletion predicted by the 10 Myr isochrone.
Our data appear inconsistent
with an age spread of this magnitude.

Whilst it is tempting to consider the link between EW[Li], luminosity
and radius as confirmation of an age spread, it is evident from Figure 2
that cold accretion can `accelerate' the evolution of a star
making it appear in all observable ways older. Thus the EW[Li],
luminosity and radius trends are the expected outcomes of either an
age spread or cold accretion; our data do not allow us to differentiate.

\subsection{Depleted lithium members}

It should be noted that not all cold accretion scenarios will show
detectable lithium depletion in young stars in the mass 
($0.3\,<
M_{\rm{\star}}\,<\,1.9\,\rm{M}_{\rm{\odot}}$) and age
range ($2\--6\,\rm{Myr}$) that we consider in this paper. Figure 10 shows a range of accreting
models from \cite{2010A&A...521A..44B} with a selection of
starting and accretion parameters that are summarised in Table 3.
Our data place limits on the levels of lithium depletion seen in our
sample. We observe up to $\sim10$ per cent depletion in
EW[Li] which corresponds to a factor of $\lesssim\,2$ in lithium
abundance for stars within the colour range defined by the forbidden zone.
Thus, our observations exclude the possibility
that our objects have experienced past accretion with $\dot{M}\,\geq\,5\times10^{-4}\,\rm{M}_{\odot}\,\rm{yr}^{-1}$. The only exception might be in the case of an extremely high-mass
  initial core ($M_{\rm{init}}>\,0.03\,\rm{M}_{\odot}$). An initial core
  mass $>\,0.03\,\rm{M}_{\odot} $ is at least an order of magnitude larger than expected from calculations
  \citep[see][and references therein]{2012ApJ...756..118B} of collapse
  from first to second Larson core \citep{1969MNRAS.145..271L}. Also, these high initial mass models do not explain the full range of
luminosities seen in the CMD and thus do not help to explain the
observed luminosity dispersion. \\

\begin{figure}
\begin{center}  
\includegraphics[width=\columnwidth]{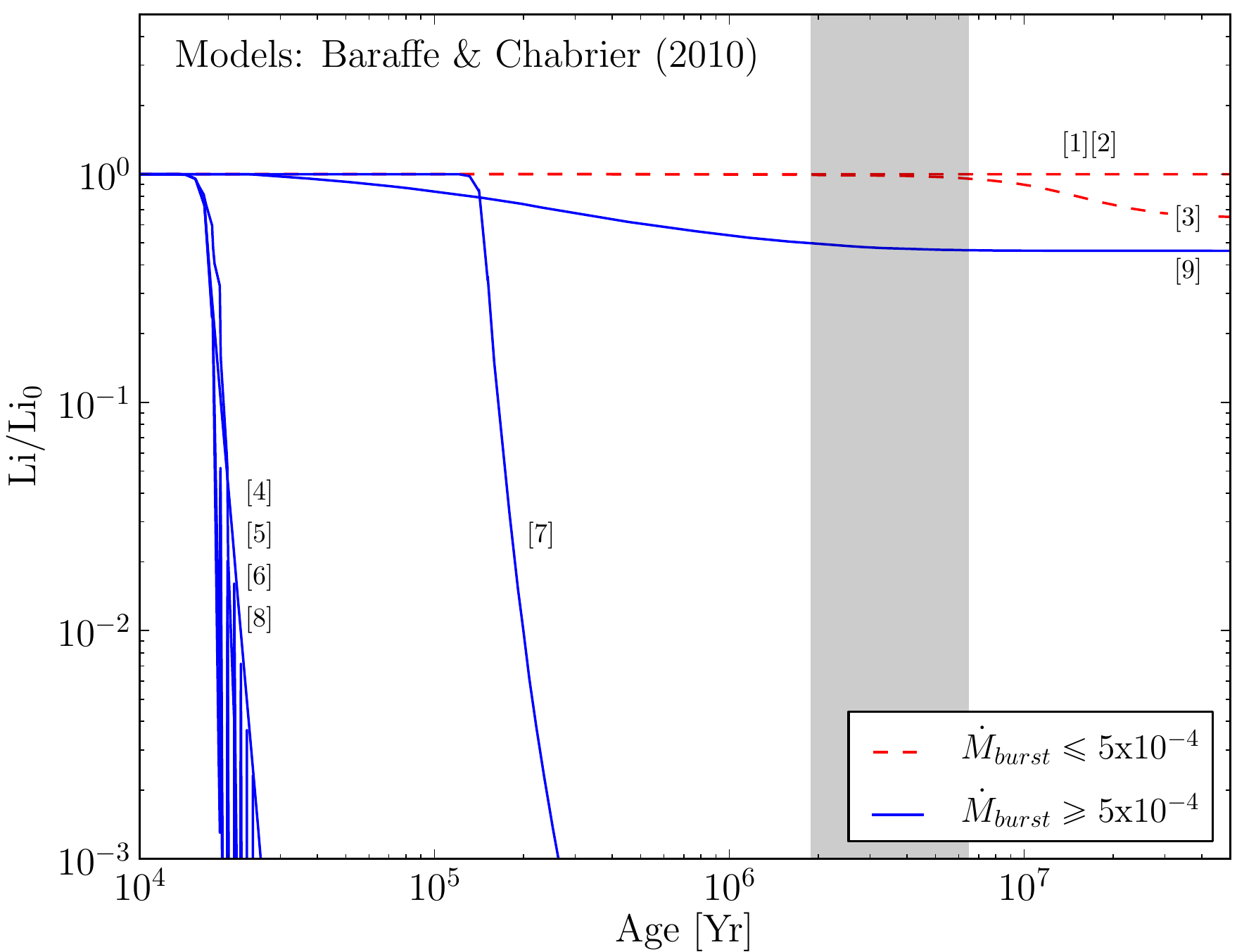}
\end{center}
\caption{Evolution of the surface lithium abundance divided by initial
Li abundance for a selection of models producing stars in the range
$0.7\--1\,\rm{M}_{\rm{\odot}}$
\protect\citep[from][]{2010A&A...521A..44B}.
Solid lines (blue) are
those using accretion burst rates $\dot{M} = 5 \times
10^{-4} \rm{M}_{\rm{\odot}} \rm{yr}^{-1}$. Dashed (red) lines are those using $\dot{M} = 1 \times
10^{-4} \rm{M}_{\rm{\odot}} \rm{yr}^{-1}$. For model details see Table 3. Shaded
grey region encompasses age estimates for the
clusters in this study.}
\end{figure}

\begin{table*}
\begin{center}
\caption{Parameters for accreting models plotted in Figure 10.}
\begin{tabular}{ccccccc}
\hline
Ref&$M_{\rm{init}}$&$\dot{M}$&No.&$T_{\rm{quiet}}$&Time to complete &Final mass\\
&(M$_{\odot}$)&(M$_{\odot}
\rm{yr}^{-1}$)&bursts&(yr)&accretion (yr) &(M$_{\odot}$)\\
\hline
1&0.001&$1\times10^{-4}$&100&$10^{3}$&$1.1\times10^{5}$&1.0\\
2&0.01&$1\times10^{-4}$&99&$10^{3}$&$1.1\times10^{5}$&1.0\\
3&0.01&$1\times10^{-4}$&69&$10^{3}$&$7.6\times10^{4}$&0.7\\
4&0.01&$5\times10^{-4}$ &14&$10^{3}$&$1.5\times10^{4}$&0.7\\
5&0.01&$5\times10^{-4}$ &16&$10^{3}$&$1.8\times10^{4}$&0.8\\
6&0.01&$5\times10^{-4}$&20&$10^{3}$&$2.2\times10^{4}$&1.0\\
7&0.01&$5\times10^{-4}$ &14&$10^{4}$&$1.4\times10^{5}$&0.7\\
8&0.02&$5\times10^{-4}$ &20&$10^{3}$&$2.2\times10^{4}$&1.0\\
9&0.03&$5\times10^{-4}$&20&$10^{3}$&$2.2\times10^{4}$&1.0\\
\hline

\end{tabular}
\end{center}
\end{table*}

The lack of strongly lithium depleted members in our sample is
consistent with the findings
of \cite{2005ApJ...626L..49P} for the ONC in our mass range, and with the only available literature sample 
for NGC 2264 \citep{1998AJ....116..254K}.
 The \cite{2005ApJ...626L..49P} sample has 82
stars in the ONC, in the mass range
$0.4\--1\,\rm{M}_{\odot}$. Significant lithium
depletion is cited for 4 of these stars, however (as may be seen in Figure 7)
these depleted objects are found to be broadly consistent with the
spread in EW[Li] seen in our
data. The levels of depletion identified are far less than we expect
for cold accretion at high rates and no stars are shown to fall within
the forbidden zone.

\cite{1998AJ....116..254K}  finds that lithium abundance is indistinguishable from meteoritic values in a sample of
12 stars of spectral type range G0V to K3V in NGC 2264. The 3 stars
closest in mass to our sample (K3V) exhibit EW[Li]
ranging from 0.49 - 0.75 \AA, which falls firmly within the range
observed in our data. 

Below the mass range that we explore, \cite{2007ApJ...659L..41P}
present lithium measurements for 4 objects of
$M_{\star}\,<\,0.3\,\rm M_{\odot}$. Their EW[Li] measurements range between $0.3 - 0.44\,\rm{\AA}$,
implying lithium depletion of 1 to 2 orders of magnitude, and ages of
15-30 Myr according to non-accreting models. Objects as old as this in
the mass range we explore would exhibit levels of lithium far below
that which we observe.

\subsection{Isochronally `old' stars}

In Section 2.2 we described the selection of seven stars that appear
faint with respect to the ONC cluster CMD and referred to these as the
LLG. These stars could be interpreted as being
older than the bulk of the population if their CMD position was
considered a reliable diagnostic of age. However, four of these stars were shown in Section 4.1
to be non-members based on their RV. Of the remaining
three, two lie below the 10 Myr isochrone and one lies below the
100 Myr isochrone.
Whilst in CMD space these objects
appear old, their measured EW[Li] and veiling ratio are entirely
consistent with the other apparently younger measured stars in the
cluster. If the
apparently oldest star has an age that is genuinely $>\,$100 Myr, then we would expect
its EW[Li] to have depleted by approximately two thirds, which we do
not see. Thus the lithium age and CMD age are in conflict. We now
discuss these three stars, reconcile their photometric and EW[Li]
properties and show that none are likely to be genuinely `old' compared with the
rest of the cluster.\\

\begin{description}
\item{\bf H676}. In CMD space, this star is apparently far older than the rest of the cluster with an
isochronal age of $>\,$100 Myr. $Hubble$ Space Telescope (HST) images
\citep{2005AJ....129..382S} reveal an edge-on massive
flared disc and accompanying Herbig-Haro jet. Its exhibited
low luminosity appears to be a function of disc extinction due to the
geometry of the system. It is clear
from studies by
\cite{2004MNRAS.351..607W} and \cite{,2010MNRAS.409.1307M} that 
circumstellar disc and accretion structures associated with young pre-MS stars can have a significant
impact on the position of an object in CMD space. The disc is a clear indicator of youth and
thus the age of the object should be considered consistent with its
EW[Li] age of $<\,$10 Myr.\\

\item{\bf H198}. This object sits below the $10\,\rm{Myr}$ isochrone
  in the CMD and yet has an unveiled EW[Li] of $0.57\pm 0.03\,$\AA, which is
  consistent with a star that is
  undepleted in lithium. HST imagery
  \citep{2000AJ....119.2919B} 
  reveals a resolved proto-planetary
  disc, visible in H$\alpha$ emission. It also
  exhibits strong evidence of interaction with a fast wind and strong
  UV flux from the star $\Theta^{\rm{1}}\,$Ori C. This is evident from
  the strong bow shock where the T-Tauri wind interacts with the wind
  from the O-star. The intense UV irradiation may cause the disc to
be heated, increase in thickness and photo-ablate \citep{1998ApJ...499..758J}.
An inflated disc and downstream flow of dusty ablating material may be
obscuring the central star and decreasing
its apparent luminosity.\\

\item{\bf H6}. This object also sits below the $10\,\rm{Myr}$ isochrone
  in the CMD and has an unveiled EW[Li] which is
  consistent with a star that is
  undepleted in lithium. It is found to be highly extincted in
  photometry by both \cite{1997AJ....113.1733H} and
  \cite{2009ApJS..183..261D}. $A_{V}$ values are 3.96
  and 5.30 respectively, measured at the two epochs which are separated by over a
  decade. Infrared colour indices ($J-H$) = 1.32
  and ($J-K$) = 2.12 also indicate strong reddening, suggesting that this object is deeply embedded within
  its envelope or within the OMC1 molecular cloud.  When subject to
  dereddening and extinction correction by \cite{1997AJ....113.1733H},
  it appears that the luminosity may be underestimated.
Dereddening vectors typically
lie almost parallel to the pre-MS in $M_{V}$ vs. $(V-I_{\rm{c}})$
if the ratio of total to selective extinction is the typical
interstellar value of $R_{V} \sim 3.2$
\citep{1998A&A...333..231B}. \cite{1990ASPC...12...63M} finds though
that $R_{V}$ in lines of sight through `outer cloud dust' can be up to
$R_{V}$ = 6. A higher value of $R_{V}$ increases the gradient of
the reddening vector in $M_{V}$ vs. $(V-I _{\rm{c}})_{\rm{0}}$
and leads us to infer an
intrinsically brighter dereddened star than that represented in the CMD.\\
\end{description}

\section{Conclusions}

We have performed a spectral analysis of two young clusters, NGC 2264
and the Orion Nebula Cluster around the 6708\AA\ lithium doublet at a
resolution of $R \sim10\,000$. Correction has been made for accretion
veiling, yielding estimates of the EW of
the lithium line for 94 stars (76 single and 18 binary) in the ONC
field and 74 stars (54 single and 20 binary) in the NGC 2264 field. Comparison has been made between measured EW[Li] and theory. 
We draw the following conclusions from our analysis.

\begin{enumerate}
\item No evidence is found in 168 stars (130 unveiled stars and 38 spectroscopic
  binaries with lower limits) for objects with EW[Li] $<0.4\,\rm{\AA}$. This finding is
  consistent with that of \cite{1998AJ....116..254K} for 12 stars in
  NGC 2264 and 82 stars in the ONC of
  \cite{2005ApJ...626L..49P}, but we have now
  almost tripled the available sample. The 262 stars now observed in
  the two clusters place limits
  on cold accretion models, indicating that burst accretion
  rates of $\dot{M}\ge5\times10^{-4}\,\rm{M}_{\odot}\,\rm{yr}^{-1}$ occur in less than
  0.5 per cent of $0.3 \le M_{\star} \le 1.9\,\rm{M}_{\odot}$
  stars. This mass accretion rate can only be exceeded without
  depleting lithium if the initial core mass is $>\,0.03\,\rm{M}_{\odot}$,
  however such models seem incapable of producing the luminosity
  dispersion seen in CMDs. This initial core mass is also at
  least an order of magnitude larger than predicted by theory
  \citep[see][and references therein]{2012ApJ...756..118B}.\\

\item Approximately 10 per cent depletion in EW[Li] is seen between the
  brighter and fainter populations in both clusters and the depletion appears to be
  broadly independent of mass. This could be due to
an age spread if the convective mixing length parameter ($\alpha$) is
close to 1.9. The scatter in
age implied by the scatter in EW[Li] is similar to the small age spread
implied by the scatter in CMD position. If $\alpha$ is closer to 1.0
then a large age spread ($>10$ Myr) is inconsistent with our observations and another
explanation must be sought. The spread in EW[Li] is also consistent
with the accelerated evolution of stellar radius and lithium abundance
described by models of past cold accretion at rates of $\dot{M}<5\times 10^{-4}$M$_{\rm{\odot}}$ yr$^{-1}$. We cannot tell which of the above scenarios are
preferred as the data and models do not allow us to distinguish between
the age spread and accretion mechanisms.\\

\item We targeted seven previously identified members in the ONC whose
  position in the CMD indicated they may be old. Four 
turned out to be radial velocity non-members. The other three objects
exhibit EW[Li] consistent with very young stars but are also found to be
  directly associated with circumstellar discs or strong
  extinction. Whilst in each case we cannot be sure that these
  features are the reason for the discrepant luminosity, it adds
  weight to the idea that discs and extinction contribute to the
  observed luminosity dispersion in pre-MS CMDs. No correlation is apparent between extinction corrected
  V magnitudes and derived $A_{V}$ across the population as a whole.\\ 

\item The median accretion veiling found for single stars in the ONC ($r_{j}$ = 0.24) is greater than
  than seen in NGC 2264 ($r_{j}$ = 0.13). In addition, the ONC
  displays 9 highly veiled objects
  ($r_{j} > 1$) whereas none are seen in NGC 2264. The highest
  veiling ratio seen in the ONC is 4.0 and in NGC 2264 is 1.0.\\

\item When establishing membership lists for young clusters ($<10\,\rm{Myr}$), lithium detection is a
  necessary requisite for membership confirmation in the mass range $0.3
  \le M_{\star} \le 1.9\,\rm{M}_{\odot}$. If lithium is not
  detected, a star must have less than 0.5 per cent chance of
  being a young cluster star.

\end{enumerate}

\section*{Acknowledgments}

DJS is funded by a UK Science and Technology Facilities Council (STFC)
studentship. The authors wish to thank Isabelle Baraffe for providing
cold accretion models and useful discussions. Spectra were extracted and calibrated using the AF2 pipeline developed by Richard
Jackson. This research is based on observations made with the
\textit{William Herschel} Telescope operated on the island of La Palma
by the \textit{Isaac Newton} Group (ING) in the Spanish Observatorio del Roque de los Muchachos of the Instituto de Astrofisica de Canarias. This research has made use of archival data products from the Two-Micron All-Sky Survey (2MASS), which is a joint project of the University of Massachusetts and the Infrared Processing and Analysis Center, funded by the National Aeronautics and Space Administration (NASA) and the National Science Foundation.

\bibliographystyle{mn2e}
\bibliography{Li_paper_arxiv}

\begin{thebibliography}{}

\bibitem[\protect\citeauthoryear{{Anders} \& {Grevesse}}{{Anders} \&
  {Grevesse}}{1989}]{1989GeCoA..53..197A}
{Anders} E.,  {Grevesse} N.,  1989, \gca, 53, 197

\bibitem[\protect\citeauthoryear{{Ballesteros-Paredes} \&
  {Hartmann}}{{Ballesteros-Paredes} \& {Hartmann}}{2007}]{2007RMxAA..43..123B}
{Ballesteros-Paredes} J.,  {Hartmann} L.,  2007, \rmxaa, 43, 123

\bibitem[\protect\citeauthoryear{{Bally}, {O'Dell} \& {McCaughrean}}{{Bally}
  et~al.}{2000}]{2000AJ....119.2919B}
{Bally} J.,  {O'Dell} C.~R.,    {McCaughrean} M.~J.,  2000, \aj, 119, 2919

\bibitem[\protect\citeauthoryear{{Baraffe} \& {Chabrier}}{{Baraffe} \&
  {Chabrier}}{2010}]{2010A&A...521A..44B}
{Baraffe} I.,  {Chabrier} G.,  2010, \aap, 521, A44

\bibitem[\protect\citeauthoryear{{Baraffe}, {Chabrier}, {Allard} \&
  {Hauschildt}}{{Baraffe} et~al.}{2002}]{2002A&A...382..563B}
{Baraffe} I.,  {Chabrier} G.,  {Allard} F.,    {Hauschildt} P.~H.,  2002, \aap,
  382, 563

\bibitem[\protect\citeauthoryear{{Baraffe}, {Chabrier} \& {Gallardo}}{{Baraffe}
  et~al.}{2009}]{2009ApJ...702L..27B}
{Baraffe} I.,  {Chabrier} G.,    {Gallardo} J.,  2009, \apjl, 702, L27

\bibitem[\protect\citeauthoryear{{Baraffe}, {Vorobyov} \& {Chabrier}}{{Baraffe}
  et~al.}{2012}]{2012ApJ...756..118B}
{Baraffe} I.,  {Vorobyov} E.,    {Chabrier} G.,  2012, \apj, 756, 118

\bibitem[\protect\citeauthoryear{{Baxter}, {Covey}, {Muench}, {F{\H
  u}r{\'e}sz}, {Rebull} \& {Szentgyorgyi}}{{Baxter}
  et~al.}{2009}]{2009AJ....138..963B}
{Baxter} E.~J.,  {Covey} K.~R.,  {Muench} A.~A.,  {F{\H u}r{\'e}sz} G.,
  {Rebull} L.,    {Szentgyorgyi} A.~H.,  2009, \aj, 138, 963

\bibitem[\protect\citeauthoryear{{Bell}, {Naylor}, {Mayne}, {Jeffries} \&
  {Littlefair}}{{Bell} et~al.}{2013}]{2013MNRAS.428.3178B}
{Bell} C.~P.~M.,  {Naylor} T.,  {Mayne} N.~J.,  {Jeffries} R.~D.,
  {Littlefair} S.~P.,  2013, \mnrassub

\bibitem[\protect\citeauthoryear{{Bessell}, {Castelli} \& {Plez}}{{Bessell}
  et~al.}{1998}]{1998A&A...333..231B}
{Bessell} M.~S.,  {Castelli} F.,    {Plez} B.,  1998, \aap, 333, 231

\bibitem[\protect\citeauthoryear{{Biazzo}, {Melo}, {Pasquini}, {Randich},
  {Bouvier} \& {Delfosse}}{{Biazzo} et~al.}{2009}]{2009A&A...508.1301B}
{Biazzo} K.,  {Melo} C.~H.~F.,  {Pasquini} L.,  {Randich} S.,  {Bouvier} J.,
  {Delfosse} X.,  2009, \aap, 508, 1301

\bibitem[\protect\citeauthoryear{{Carlsson}, {Rutten}, {Bruls} \&
  {Shchukina}}{{Carlsson} et~al.}{1994}]{1994A&A...288..860C}
{Carlsson} M.,  {Rutten} R.~J.,  {Bruls} J.~H.~M.~J.,    {Shchukina} N.~G.,
  1994, \aap, 288, 860

\bibitem[\protect\citeauthoryear{{Da Rio}, {Gouliermis} \& {Gennaro}}{{Da Rio}
  et~al.}{2010}]{2010ApJ...723..166D}
{Da Rio} N.,  {Gouliermis} D.~A.,    {Gennaro} M.,  2010, \apj, 723, 166

\bibitem[\protect\citeauthoryear{{Da Rio}, {Robberto}, {Soderblom}, {Panagia},
  {Hillenbrand}, {Palla} \& {Stassun}}{{Da Rio}
  et~al.}{2009}]{2009ApJS..183..261D}
{Da Rio} N.,  {Robberto} M.,  {Soderblom} D.~R.,  {Panagia} N.,  {Hillenbrand}
  L.~A.,  {Palla} F.,    {Stassun} K.,  2009, \apjs, 183, 261

\bibitem[\protect\citeauthoryear{{Dahm} \& {Simon}}{{Dahm} \&
  {Simon}}{2005}]{2005AJ....129..829D}
{Dahm} S.~E.,  {Simon} T.,  2005, \aj, 129, 829

\bibitem[\protect\citeauthoryear{{Dodin} \& {Lamzin}}{{Dodin} \&
  {Lamzin}}{2012}]{2012arXiv1209.1851D}
{Dodin} A.~V.,  {Lamzin} S.~A.,  2012, ArXiv e-prints

\bibitem[\protect\citeauthoryear{{Elmegreen}}{{Elmegreen}}{2000}]{2000ApJ...530..277E}
{Elmegreen} B.~G.,  2000, \apj, 530, 277

\bibitem[\protect\citeauthoryear{{Enoch}, {Evans} II, {Sargent} \&
  {Glenn}}{{Enoch} et~al.}{2009}]{2009ApJ...692..973E}
{Enoch} M.~L.,  {Evans} II N.~J.,  {Sargent} A.~I.,    {Glenn} J.,  2009, \apj,
  692, 973

\bibitem[\protect\citeauthoryear{{F{\H u}r{\'e}sz}, {Hartmann}, {Megeath},
  {Szentgyorgyi} \& {Hamden}}{{F{\H u}r{\'e}sz}
  et~al.}{2008}]{2008ApJ...676.1109F}
{F{\H u}r{\'e}sz} G.,  {Hartmann} L.~W.,  {Megeath} S.~T.,  {Szentgyorgyi}
  A.~H.,    {Hamden} E.~T.,  2008, \apj, 676, 1109

\bibitem[\protect\citeauthoryear{{F{\H u}r{\'e}sz}, {Hartmann}, {Szentgyorgyi},
  {Ridge}, {Rebull}, {Stauffer}, {Latham}, {Conroy}, {Fabricant} \&
  {Roll}}{{F{\H u}r{\'e}sz} et~al.}{2006}]{2006ApJ...648.1090F}
{F{\H u}r{\'e}sz} G.,  {Hartmann} L.~W.,  {Szentgyorgyi} A.~H.,  {Ridge} N.~A.,
   {Rebull} L.,  {Stauffer} J.,  {Latham} D.~W.,  {Conroy} M.~A.,  {Fabricant}
  D.~G.,    {Roll} J.,  2006, \apj, 648, 1090

\bibitem[\protect\citeauthoryear{{Flaccomio}, {Damiani}, {Micela}, {Sciortino},
  {Harnden} Jr., {Murray} \& {Wolk}}{{Flaccomio}
  et~al.}{2002}]{2002ASPC..277..155F}
{Flaccomio} E.,  {Damiani} F.,  {Micela} G.,  {Sciortino} S.,  {Harnden} Jr.
  F.~R.,  {Murray} S.~S.,    {Wolk} S.~J.,  2002, in {Favata} F.,  {Drake}
  J.~J.,  eds, Stellar Coronae in the Chandra and XMM-NEWTON Era Vol.~277 of
  Astronomical Society of the Pacific Conference Series, {Chandra X-ray
  Observation of the Orion Nebula Cluster}.
p.~155

\bibitem[\protect\citeauthoryear{{Flaccomio}, {Micela}, {Sciortino}, {Favata},
  {Corbally} \& {Tomaney}}{{Flaccomio} et~al.}{1999}]{1999A&A...345..521F}
{Flaccomio} E.,  {Micela} G.,  {Sciortino} S.,  {Favata} F.,  {Corbally} C.,
  {Tomaney} A.,  1999, \aap, 345, 521

\bibitem[\protect\citeauthoryear{{Gahm}, {Walter}, {Stempels}, {Petrov} \&
  {Herczeg}}{{Gahm} et~al.}{2008}]{2008A&A...482L..35G}
{Gahm} G.~F.,  {Walter} F.~M.,  {Stempels} H.~C.,  {Petrov} P.~P.,    {Herczeg}
  G.~J.,  2008, \aap, 482, L35

\bibitem[\protect\citeauthoryear{{Guenther} \& {Hessman}}{{Guenther} \&
  {Hessman}}{1994}]{1994ASPC...62..132G}
{Guenther} E.,  {Hessman} F.~V.,  1994, in {The} P.~S.,  {Perez} M.~R.,   {van
  den Heuvel} E.~P.~J.,  eds, The Nature and Evolutionary Status of Herbig
  Ae/Be Stars Vol.~62 of Astronomical Society of the Pacific Conference Series,
  {The veiling continuum of T Tauri stars}.
p.~132

\bibitem[\protect\citeauthoryear{{Hartigan}, {Hartmann}, {Kenyon}, {Hewett} \&
  {Stauffer}}{{Hartigan} et~al.}{1989}]{1989ApJS...70..899H}
{Hartigan} P.,  {Hartmann} L.,  {Kenyon} S.,  {Hewett} R.,    {Stauffer} J.,
  1989, \apjs, 70, 899

\bibitem[\protect\citeauthoryear{{Hartmann}, {Cassen} \& {Kenyon}}{{Hartmann}
  et~al.}{1997}]{1997ApJ...475..770H}
{Hartmann} L.,  {Cassen} P.,    {Kenyon} S.~J.,  1997, \apj, 475, 770

\bibitem[\protect\citeauthoryear{{Hartmann}, {Zhu} \& {Calvet}}{{Hartmann}
  et~al.}{2011}]{2011arXiv1106.3343H}
{Hartmann} L.,  {Zhu} Z.,    {Calvet} N.,  2011, ArXiv e-prints

\bibitem[\protect\citeauthoryear{{Herbst}, {Bailer-Jones}, {Mundt},
  {Meisenheimer} \& {Wackermann}}{{Herbst} et~al.}{2002}]{2002A&A...396..513H}
{Herbst} W.,  {Bailer-Jones} C.~A.~L.,  {Mundt} R.,  {Meisenheimer} K.,
  {Wackermann} R.,  2002, \aap, 396, 513

\bibitem[\protect\citeauthoryear{{Hillenbrand}}{{Hillenbrand}}{1997}]{1997AJ....113.1733H}
{Hillenbrand} L.~A.,  1997, \aj, 113, 1733

\bibitem[\protect\citeauthoryear{{Horne}}{{Horne}}{1986}]{1986PASP...98..609H}
{Horne} K.,  1986, \pasp, 98, 609

\bibitem[\protect\citeauthoryear{{Hosokawa}, {Offner} \& {Krumholz}}{{Hosokawa}
  et~al.}{2011}]{2011ApJ...738..140H}
{Hosokawa} T.,  {Offner} S.~S.~R.,    {Krumholz} M.~R.,  2011, \apj, 738, 140

\bibitem[\protect\citeauthoryear{{Jeffries}}{{Jeffries}}{2007}]{2007MNRAS.381.1169J}
{Jeffries} R.~D.,  2007, \mnras, 381, 1169

\bibitem[\protect\citeauthoryear{{Jeffries}, {Littlefair}, {Naylor} \&
  {Mayne}}{{Jeffries} et~al.}{2011}]{2011MNRAS.418.1948J}
{Jeffries} R.~D.,  {Littlefair} S.~P.,  {Naylor} T.,    {Mayne} N.~J.,  2011,
  \mnras, 418, 1948

\bibitem[\protect\citeauthoryear{{Jeffries} \& {Oliveira}}{{Jeffries} \&
  {Oliveira}}{2005}]{2005MNRAS.358...13J}
{Jeffries} R.~D.,  {Oliveira} J.~M.,  2005, \mnras, 358, 13

\bibitem[\protect\citeauthoryear{{Jeffries}, {Oliveira}, {Barrado y
  Navascu{\'e}s} \& {Stauffer}}{{Jeffries} et~al.}{2003}]{2003MNRAS.343.1271J}
{Jeffries} R.~D.,  {Oliveira} J.~M.,  {Barrado y Navascu{\'e}s} D.,
  {Stauffer} J.~R.,  2003, \mnras, 343, 1271

\bibitem[\protect\citeauthoryear{{Johnstone}, {Hollenbach} \&
  {Bally}}{{Johnstone} et~al.}{1998}]{1998ApJ...499..758J}
{Johnstone} D.,  {Hollenbach} D.,    {Bally} J.,  1998, \apj, 499, 758

\bibitem[\protect\citeauthoryear{{Kenyon} \& {Hartmann}}{{Kenyon} \&
  {Hartmann}}{1995}]{1995ApJS..101..117K}
{Kenyon} S.~J.,  {Hartmann} L.,  1995, \apjs, 101, 117

\bibitem[\protect\citeauthoryear{{Kenyon}, {Hartmann}, {Strom} \&
  {Strom}}{{Kenyon} et~al.}{1990}]{1990AJ.....99..869K}
{Kenyon} S.~J.,  {Hartmann} L.~W.,  {Strom} K.~M.,    {Strom} S.~E.,  1990,
  \aj, 99, 869

\bibitem[\protect\citeauthoryear{{King}}{{King}}{1998}]{1998AJ....116..254K}
{King} J.~R.,  1998, \aj, 116, 254

\bibitem[\protect\citeauthoryear{{Lamm}, {Bailer-Jones}, {Mundt}, {Herbst} \&
  {Scholz}}{{Lamm} et~al.}{2004}]{2004A&A...417..557L}
{Lamm} M.~H.,  {Bailer-Jones} C.~A.~L.,  {Mundt} R.,  {Herbst} W.,    {Scholz}
  A.,  2004, \aap, 417, 557

\bibitem[\protect\citeauthoryear{{Larson}}{{Larson}}{1969}]{1969MNRAS.145..271L}
{Larson} R.~B.,  1969, \mnras, 145, 271

\bibitem[\protect\citeauthoryear{{Littlefair}, {Naylor}, {Mayne}, {Saunders} \&
  {Jeffries}}{{Littlefair} et~al.}{2011}]{2011MNRAS.413L..56L}
{Littlefair} S.~P.,  {Naylor} T.,  {Mayne} N.~J.,  {Saunders} E.,    {Jeffries}
  R.~D.,  2011, \mnras, 413, L56

\bibitem[\protect\citeauthoryear{{Mathis}}{{Mathis}}{1990}]{1990ASPC...12...63M}
{Mathis} J.~S.,  1990, in {Blitz} L.,  ed., The Evolution of the Interstellar
  Medium Vol.~12 of Astronomical Society of the Pacific Conference Series,
  {Interstellar dust and extinction}.
pp 63--77

\bibitem[\protect\citeauthoryear{{Mayne} \& {Harries}}{{Mayne} \&
  {Harries}}{2010}]{2010MNRAS.409.1307M}
{Mayne} N.~J.,  {Harries} T.~J.,  2010, \mnras, 409, 1307

\bibitem[\protect\citeauthoryear{{Mayne} \& {Naylor}}{{Mayne} \&
  {Naylor}}{2008}]{2008MNRAS.386..261M}
{Mayne} N.~J.,  {Naylor} T.,  2008, \mnras, 386, 261

\bibitem[\protect\citeauthoryear{{Mayne}, {Naylor}, {Littlefair}, {Saunders} \&
  {Jeffries}}{{Mayne} et~al.}{2007}]{2007MNRAS.375.1220M}
{Mayne} N.~J.,  {Naylor} T.,  {Littlefair} S.~P.,  {Saunders} E.~S.,
  {Jeffries} R.~D.,  2007, \mnras, 375, 1220

\bibitem[\protect\citeauthoryear{{McNamara}}{{McNamara}}{1976}]{1976AJ.....81..845M}
{McNamara} B.~J.,  1976, \aj, 81, 845

\bibitem[\protect\citeauthoryear{{Menten}, {Reid}, {Forbrich} \&
  {Brunthaler}}{{Menten} et~al.}{2007}]{2007AA...474..515M}
{Menten} K.~M.,  {Reid} M.~J.,  {Forbrich} J.,    {Brunthaler} A.,  2007, \aap,
  474, 515

\bibitem[\protect\citeauthoryear{{Naylor}}{{Naylor}}{2009}]{2009MNRAS.399..432N}
{Naylor} T.,  2009, \mnras, 399, 432

\bibitem[\protect\citeauthoryear{{Palla}, {Randich}, {Flaccomio} \&
  {Pallavicini}}{{Palla} et~al.}{2005}]{2005ApJ...626L..49P}
{Palla} F.,  {Randich} S.,  {Flaccomio} E.,    {Pallavicini} R.,  2005, \apjl,
  626, L49

\bibitem[\protect\citeauthoryear{{Palla}, {Randich}, {Pavlenko}, {Flaccomio} \&
  {Pallavicini}}{{Palla} et~al.}{2007}]{2007ApJ...659L..41P}
{Palla} F.,  {Randich} S.,  {Pavlenko} Y.~V.,  {Flaccomio} E.,    {Pallavicini}
  R.,  2007, \apjl, 659, L41

\bibitem[\protect\citeauthoryear{{Palla} \& {Stahler}}{{Palla} \&
  {Stahler}}{2000}]{2000ApJ...540..255P}
{Palla} F.,  {Stahler} S.~W.,  2000, \apj, 540, 255

\bibitem[\protect\citeauthoryear{{Pflamm-Altenburg} \&
  {Kroupa}}{{Pflamm-Altenburg} \& {Kroupa}}{2007}]{2007MNRAS.375..855P}
{Pflamm-Altenburg} J.,  {Kroupa} P.,  2007, \mnras, 375, 855

\bibitem[\protect\citeauthoryear{{Pinsonneault}}{{Pinsonneault}}{1997}]{1997ARA&A..35..557P}
{Pinsonneault} M.,  1997, \araa, 35, 557

\bibitem[\protect\citeauthoryear{{Rhode}, {Herbst} \& {Mathieu}}{{Rhode}
  et~al.}{2001}]{2001AJ....122.3258R}
{Rhode} K.~L.,  {Herbst} W.,    {Mathieu} R.~D.,  2001, \aj, 122, 3258

\bibitem[\protect\citeauthoryear{{Sacco}, {Randich}, {Franciosini},
  {Pallavicini} \& {Palla}}{{Sacco} et~al.}{2007}]{2007A&A...462L..23S}
{Sacco} G.~G.,  {Randich} S.,  {Franciosini} E.,  {Pallavicini} R.,    {Palla}
  F.,  2007, \aap, 462, L23

\bibitem[\protect\citeauthoryear{{Siess}, {Dufour} \& {Forestini}}{{Siess}
  et~al.}{2000}]{2000A&A...358..593S}
{Siess} L.,  {Dufour} E.,    {Forestini} M.,  2000, \aap, 358, 593

\bibitem[\protect\citeauthoryear{{Smith}, {Bally}, {Licht} \&
  {Walawender}}{{Smith} et~al.}{2005}]{2005AJ....129..382S}
{Smith} N.,  {Bally} J.,  {Licht} D.,    {Walawender} J.,  2005, \aj, 129, 382

\bibitem[\protect\citeauthoryear{{Tan}, {Krumholz} \& {McKee}}{{Tan}
  et~al.}{2006}]{2006ApJ...641L.121T}
{Tan} J.~C.,  {Krumholz} M.~R.,    {McKee} C.~F.,  2006, \apjl, 641, L121

\bibitem[\protect\citeauthoryear{{Tout}, {Livio} \& {Bonnell}}{{Tout}
  et~al.}{1999}]{1999MNRAS.310..360T}
{Tout} C.~A.,  {Livio} M.,    {Bonnell} I.~A.,  1999, \mnras, 310, 360

\bibitem[\protect\citeauthoryear{{van Hamme}}{{van
  Hamme}}{1993}]{1993AJ....106.2096V}
{van Hamme} W.,  1993, \aj, 106, 2096

\bibitem[\protect\citeauthoryear{{Viallet} \& {Baraffe}}{{Viallet} \&
  {Baraffe}}{2012}]{2012arXiv1209.1812V}
{Viallet} M.,  {Baraffe} I.,  2012, ArXiv e-prints

\bibitem[\protect\citeauthoryear{{Vorobyov} \& {Basu}}{{Vorobyov} \&
  {Basu}}{2005}]{2005ApJ...633L.137V}
{Vorobyov} E.~I.,  {Basu} S.,  2005, \apjl, 633, L137

\bibitem[\protect\citeauthoryear{{Walker}, {Wood}, {Lada}, {Robitaille},
  {Bjorkman} \& {Whitney}}{{Walker} et~al.}{2004}]{2004MNRAS.351..607W}
{Walker} C.,  {Wood} K.,  {Lada} C.~J.,  {Robitaille} T.,  {Bjorkman} J.~E.,
  {Whitney} B.,  2004, \mnras, 351, 607

\bibitem[\protect\citeauthoryear{{Zapatero Osorio}, {B{\'e}jar}, {Pavlenko},
  {Rebolo}, {Allende Prieto}, {Mart{\'{\i}}n} \& {Garc{\'{\i}}a
  L{\'o}pez}}{{Zapatero Osorio} et~al.}{2002}]{2002A&A...384..937Z}
{Zapatero Osorio} M.~R.,  {B{\'e}jar} V.~J.~S.,  {Pavlenko} Y.,  {Rebolo} R.,
  {Allende Prieto} C.,  {Mart{\'{\i}}n} E.~L.,    {Garc{\'{\i}}a L{\'o}pez}
  R.~J.,  2002, \aap, 384, 937

\end{thebibliography}

\label{lastpage}

\end{document}